\documentclass[a4paper]{article}       

\usepackage{amsmath,mathtools,amsthm}
\usepackage{amssymb, bm}
\usepackage{a4wide,authblk}
\usepackage[mathscr]{eucal}
\usepackage{bm}
\usepackage{bbm}
\usepackage[shortlabels]{enumitem}
\usepackage{hyperref}
\usepackage{booktabs}
\usepackage[usenames,dvipsnames]{xcolor}

\usepackage{tikz}
\usepackage{graphicx, caption}
\usepackage{subcaption}

\newcommand{\prob}{\mathbb{P}}
\newcommand{\Prob}[1]{\prob\left(#1\right)}

\newcommand{\expec}{\mathbb{E}}
\newcommand{\Exp}[1]{\expec\left[#1\right]}

\newcommand{\Var}[1]{\textup{Var}\left(#1\right)}

\newcommand{\sss}[1]{\scriptscriptstyle{#1}}

\newcommand{\me}{\textup{e}}
\newcommand{\dd}{{\rm d}}

\newcommand{\op}{o_{\sss\prob}}

\allowdisplaybreaks

\graphicspath{{Figures/}}

\def\bowa{\tikz[baseline=.1ex]{
		\coordinate (A) at (0,-0.5ex);
		\fill (1.5ex,-0.5ex) circle (2pt) coordinate (B);
		\coordinate (C) at (0,2.5ex);
		\fill (1.5ex,2.5ex) circle (2pt) coordinate (D);
		\fill (0.75ex,1ex) circle (2pt) coordinate (E);
		\draw (A)--(B);
		\draw (A)--(E);
		\draw (E)--(B);
		\draw (C)--(D);
		\draw (E)--(D);
		\draw (E)--(C);
		\draw[fill=white](A) circle (2pt);
		\draw[fill=white](C) circle (2pt);		}
}
\def\bowb{\tikz[baseline=.1ex]{
		\coordinate (A) at (0,-0.5ex);
		\fill (1.5ex,-0.5ex) circle (2pt) coordinate (B);
		\coordinate (C) at (0,2.5ex);
		\fill (1.5ex,2.5ex) circle (2pt) coordinate (D);
		\coordinate (E) at (0.75ex,1ex);
		\draw (A)--(B);
		\draw (A)--(E);
		\draw (E)--(B);
		\draw (C)--(D);
		\draw (E)--(D);
		\draw (E)--(C);
		\draw[fill=white](A) circle (2pt);
		\draw[fill=white](E) circle (2pt);
		\fill(C) circle (2pt);		}
}
\def\bowc{\tikz[baseline=.1ex]{
		\coordinate (A) at (0,-0.5ex);
		\fill (1.5ex,-0.5ex) circle (2pt) coordinate (B);
		\coordinate (C) at (0,2.5ex);
		\fill (1.5ex,2.5ex) circle (2pt) coordinate (D);
		\coordinate (E) at (0.75ex,1ex);
		\draw (A)--(B);
		\draw (A)--(E);
		\draw (E)--(B);
		\draw (C)--(D);
		\draw (E)--(D);
		\draw (E)--(C);
		\fill(A) circle (2pt);
		\draw[fill=white](E) circle (2pt);
		\fill(C) circle (2pt);		}
}

\def\diamonda{\tikz[baseline=.1ex]{
		\coordinate (A) at (0,-0.5ex);
		\coordinate (B) at (3ex,-0.5ex);
		\coordinate (C) at (0,2.5ex);
		\coordinate (D) at (3ex,2.5ex) ;
		\draw (A)--(B);
		\draw (A)--(C);
		\draw (C)--(B);
		\draw (C)--(D);
		\draw (B)--(D);
		\draw[fill=white](A) circle (2pt);
		\fill(B) circle (2pt);
		\fill(C) circle (2pt);	
		\draw[fill=white](D) circle (2pt);	}
}

\def\diamondb{\tikz[baseline=.1ex]{
		\coordinate (A) at (0,-0.5ex);
		\coordinate (B) at (3ex,-0.5ex);
		\coordinate (C) at (0,2.5ex);
		\coordinate (D) at (3ex,2.5ex) ;
		\draw (A)--(B);
		\draw (A)--(C);
		\draw (C)--(B);
		\draw (C)--(D);
		\draw (B)--(D);
		\draw[fill=white](A) circle (2pt);
		\fill(B) circle (2pt);
		\draw[fill=white](C) circle (2pt);	
		\fill(D) circle (2pt);	}
}

\def\diamondc{\tikz[baseline=.1ex]{
		\coordinate (A) at (0,-0.5ex);
		\coordinate (B) at (3ex,-0.5ex);
		\coordinate (C) at (0,2.5ex);
		\coordinate (D) at (3ex,2.5ex) ;
		\draw (A)--(B);
		\draw (A)--(C);
		\draw (C)--(B);
		\draw (C)--(D);
		\draw (B)--(D);
		\fill(A) circle (2pt);
		\draw[fill=white](B) circle (2pt);
		\fill(D) circle (2pt);	
		\draw[fill=white](C) circle (2pt);	}
}

\def\diamondd{\tikz[baseline=.1ex]{
		\coordinate (A) at (0,-0.5ex);
		\coordinate (B) at (3ex,-0.5ex);
		\coordinate (C) at (0,2.5ex);
		\coordinate (D) at (3ex,2.5ex) ;
		\draw (A)--(B);
		\draw (A)--(C);
		\draw (C)--(B);
		\draw (C)--(D);
		\draw (B)--(D);
		\draw[fill=white](B) circle (2pt);
		\fill(A) circle (2pt);
		\fill(C) circle (2pt);	
		\fill(D) circle (2pt);	}
}

\def\triangd{\tikz[baseline=.1ex]{
		\coordinate (A) at (0,-0.5ex);
		\coordinate (B) at (3ex,-0.5ex);
		\coordinate (C) at (1.5ex,2ex);
		\draw (A)--(B);
		\draw (A)--(C);
		\draw (C)--(B);
		\draw[fill=white](B) circle (2pt);
		\draw[fill=white](A) circle (2pt);
		\fill(C) circle (2pt);		}
}

\definecolor{color1}{HTML}{fff0f5}%
\definecolor{color2}{rgb}{0.6314,0.8549,0.7059}%
\definecolor{colorsqrt}{HTML}{ff3799}%
\definecolor{colornm}{HTML}{ffa3c8}%
\definecolor{colorn}{HTML}{4b0082}
\definecolor{colorn1tau}{HTML}{8b0088}

\tikzstyle{S1}=[fill=colornm]
\tikzstyle{S2m}=[fill=colorn1tau]
\tikzstyle{S2}=[fill=colorn]
\tikzstyle{S3}=[fill=colorsqrt]
\tikzstyle{n1}=[fill=color1]

\begin{document}
\title{Scale-free network clustering in hyperbolic and other random graphs}
\author{Clara Stegehuis}
\author{Remco van der Hofstad}
\author{Johan S.H. van Leeuwaarden}
\affil{Department of Mathematics and Computer Science \\Eindhoven University of Technology}

\maketitle

\begin{abstract}

Random graphs with power-law degrees can model scale-free networks as sparse topologies with strong degree heterogeneity. Mathematical analysis of such random graphs proved successful in explaining scale-free network properties such as resilience, navigability and small distances. We introduce a variational principle to explain how vertices tend to cluster in triangles as a function of their degrees. 
We apply the variational principle to the hyperbolic model that quickly gains popularity as a model for scale-free networks with latent geometries and clustering. We show that clustering in the hyperbolic model is non-vanishing and self-averaging, so that a single random graph sample is a good representation in the large-network limit. We also demonstrate the variational principle for some classical random graphs including the preferential attachment model and the configuration model.  
\end{abstract}

\section{Introduction}
Scale-free networks feature in many branches of science, describing connectivity patterns between particles through large graphs with strong vertex-degree heterogeneity, often modeled as a power law that lets the proportion of vertices with $k$ neighbors scale as $k^{-\tau}$. Statistical analysis suggests that the power-law exponent $\tau$ in real-world networks often lies between 2 and 3 \cite{albert1999,faloutsos1999,jeong2000,vazquez2002}, so that the vertex degree has a finite first and infinite second moment. With such power laws, vertices of extremely high degrees (also called hubs) are likely to be present, and cause scale-free properties such as small distances~\cite{hofstad2007, newman2001}, fast information spreading~\cite{dorogovtsev2008,pastor2001,boguna2004} and the absence of percolation thresholds~\cite{janson2009b, pastor2001}. Power-law degrees and hubs also crucially influence local properties such as the abundance of certain subgraphs like triangles and cliques~\cite{ostilli2014,stegehuis2017,hofstad2017b}. 

For several decades now, scientists are building the theoretical foundation for scale-free networks, using a large variety of mathematical models and approaches. Classical models like the preferential attachment model and the hidden-variable model generate mathematically tractable random graphs with power-law degrees, and were successful in explaining some of the key empirical observations for distances and information spreading. The average distance in the scale-free random graph models, for instance, was shown to scale as $\log\log n$ in the network size $n$~\cite{dereich2012,hofstad2007,dorogovtsev2002,cohen2003}, in agreement with the small distances measured in real-world networks. 


\begin{figure*}[tb]
	\begin{subfigure}[t]{0.3\linewidth}
		\centering
		\includegraphics[width=\linewidth]{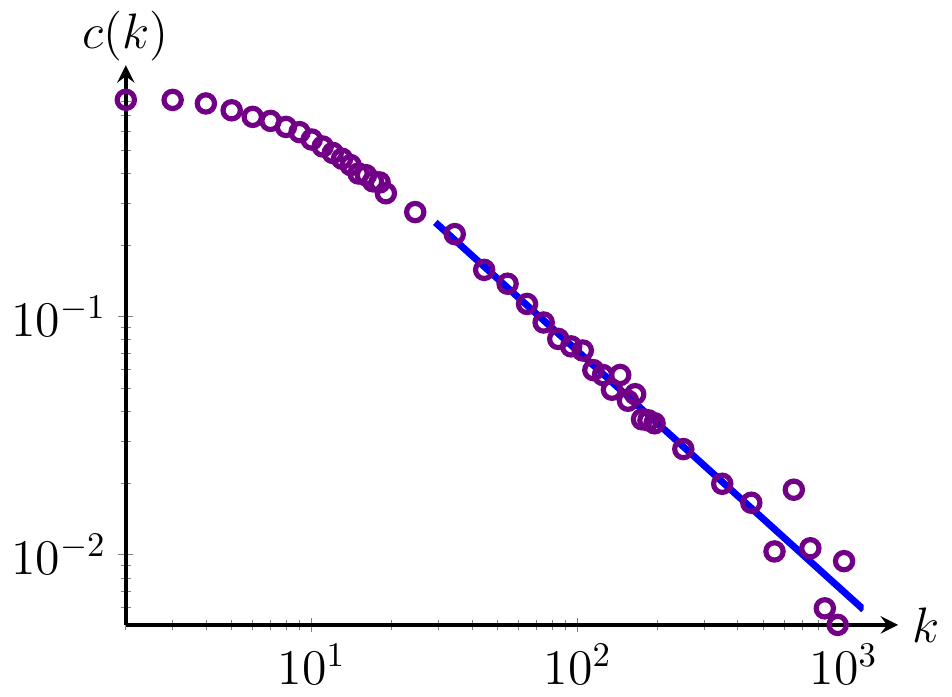}
		\caption{}
		\label{fig:chwordnet}
	\end{subfigure}
	\begin{subfigure}[t]{0.3\linewidth}
		\centering
		\includegraphics[width=\linewidth]{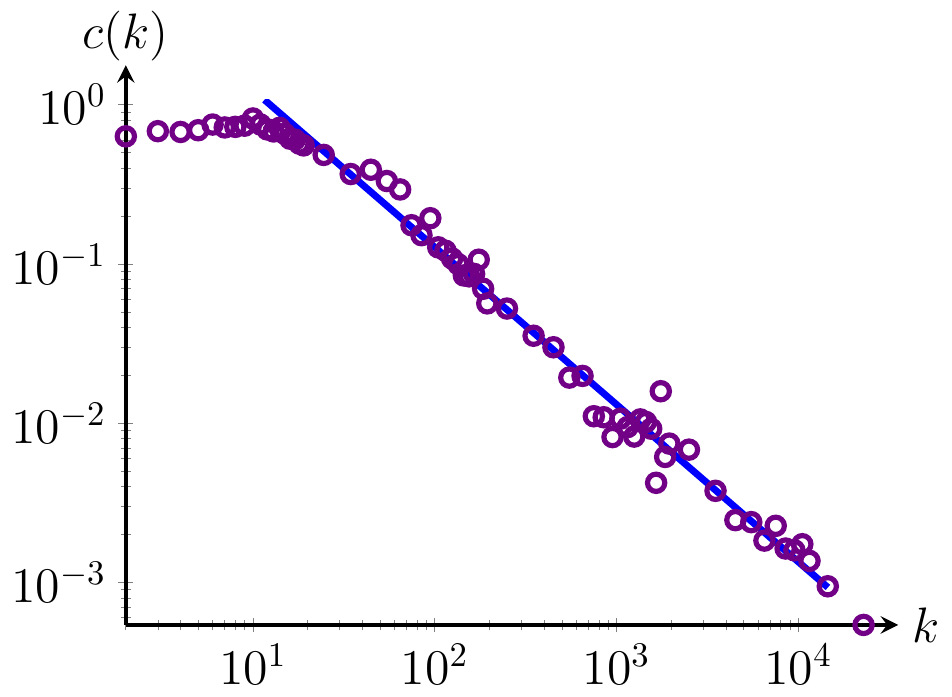}
		\caption{}
		\label{fig:chtrec}
	\end{subfigure}	
	\begin{subfigure}[t]{0.3\linewidth}
		\centering
		\includegraphics[width=\linewidth]{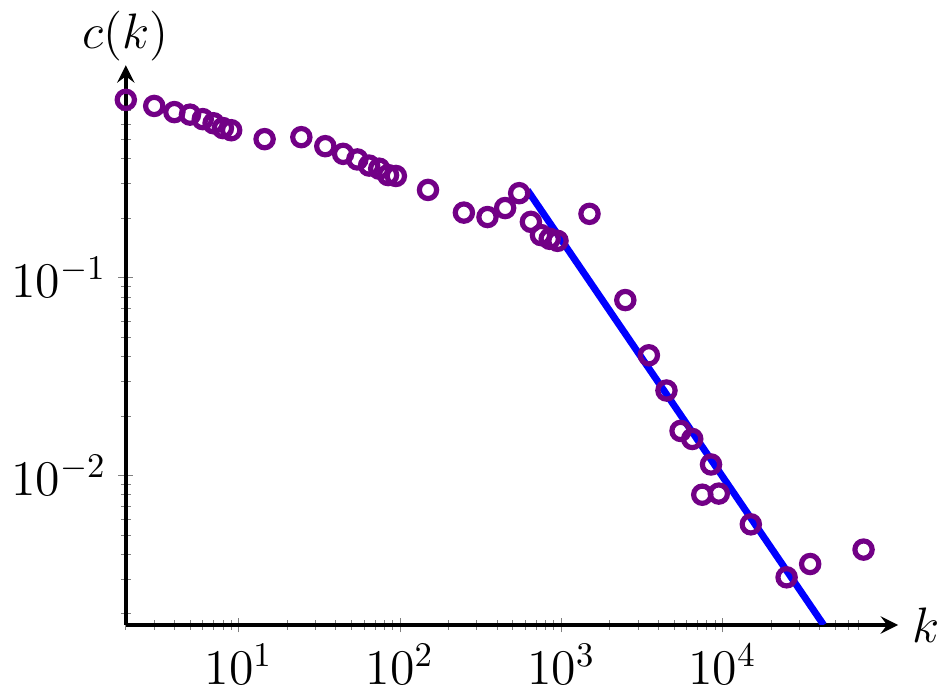}
		\caption{}
		\label{fig:chcatster}
	\end{subfigure}
	\caption{$c(k)$ for (a) the WordNet information network~\cite{miller1998}, (b) the TREC-WT10g web graph~\cite{bailey2003} and (c) the Catster/Dogster social network~\cite{konect}.}
	\label{fig:ch}
\end{figure*}

Another empirically observed scale-free property is the tendency of vertices to cluster together in groups with relatively many edges between the group members~\cite{girvan2002}. The preferential attachment model and the hidden-variable model, however, both have vanishing clustering levels when the network size grows to infinity, rendering these models unfit for modeling group formation in the large-network limit. We therefore employ the hyperbolic model, which in recent years has emerged as a new scale-free network model \cite{krioukov2010,allard2017,boguna2010,garcia-perez2016,borassi2015}. The model creates a random graph by positioning each vertex at a uniformly chosen location in the hyperbolic space, and then connecting pairs of vertices as a function of their locations. The hyperbolic model is mathematically tractable and capable of matching simultaneously the three key characteristics of real-world networks: sparseness, power-law degrees and clustering. 


The degree of clustering can be measured in terms of the local clustering coefficient $c(k)$, the probability that two neighbors of a degree-$k$ vertex are neighbors themselves, and also in terms of the average clustering coefficient $C$ that gives an overall indication of the clustering in the network. Ample empirical evidence shows that the function $c(k)$ generally decreases in $k$ according to some power law~\cite{vazquez2002,maslov2004,stegehuis2017,boguna2003,colomer2012,krioukov2012}, which suggest that the network can be viewed as a collection of subgraphs with dense connections within themselves and sparser ones between them~\cite{ravasz2003}. Randomizing real-world networks while preserving the shape of the $c(k)$-curve produces networks with very similar component sizes as well as similar hierarchical structures as the original network~\cite{colomer2013}. The shape of $c(k)$ also influences the behavior of networks under percolation~\cite{serrano2006,newman2003b}. 

Figure~\ref{fig:ch} shows the $c(k)$-curves for three different networks: an information network describing the relationships between English words (Fig.~\ref{fig:chwordnet}), a technological network describing web pages and their hyperlinks (Fig.~\ref{fig:chtrec}) and a social network (Fig.~\ref{fig:chcatster}). While these three networks are very different, their $c(k)$-curves share several similarities. First of all, $c(k)$ decays in $k$ in all three networks. Furthermore, for small values of $k$, $c(k)$ is high in all three networks, indicating the presence of non-trivial clustering. 
Taking the hyperbolic model as the network model, we obtain a precise characterization of clustering in the hyperbolic model by describing how the clustering curve $k\mapsto c(k)$ scales with $k$ and $n$. We also obtain the scaling behavior for $C$ from the results for $c(k)$. 

Studying the local clustering coefficient $c(k)$ is equivalent to studying the number of triangles where at least one of the vertices has degree $k$. We develop a novel conceptual framework, a variational principle, that finds the dominant such triangle in terms of the degrees of the other two vertices. This variational principle exploits the trade-off present in power-law networks:  high-degree vertices are well connected and therefore participate in many triangles, but high-degree vertices are rare because of the power-law degree distribution. Lower-degree vertices typically participate in fewer triangles, but occur more frequently. The variational principle finds the degrees that optimize this trade-off and reveals the structure of the three-point correlations between triplets of vertices that dictate the degree of clustering.

In Section~\ref{sec:hyperbolic} we present the variational principle and apply it to the hyperbolic model to find the typical relation between clustering, node degree and the power-law exponent of the degree distribution. In Section \ref{ssf} we present an extended version of the variational principle that can deal with general subgraphs and apply it to characterize the sample-to-sample fluctuations of clustering in the hyperbolic model. As it turns out, the  clustering curve $k\mapsto c(k)$ and the global clustering coefficient $C$ in the hyperbolic model are non-vanishing and self-averaging as $n\to\infty$. 
We then proceed to apply the variational principle to the hidden-variable model, the preferential attachment model and the random intersection graph in Section~\ref{sec:treelike}.

\section{Hyperbolic model}\label{sec:hyperbolic}

We now discuss the hyperbolic model in more detail, introduce the generic variational principle to characterize clustering, and then apply the variational principle to the hyperbolic model.

\begin{figure}[tb]
		\centering
	\includegraphics[width=0.32\textwidth]{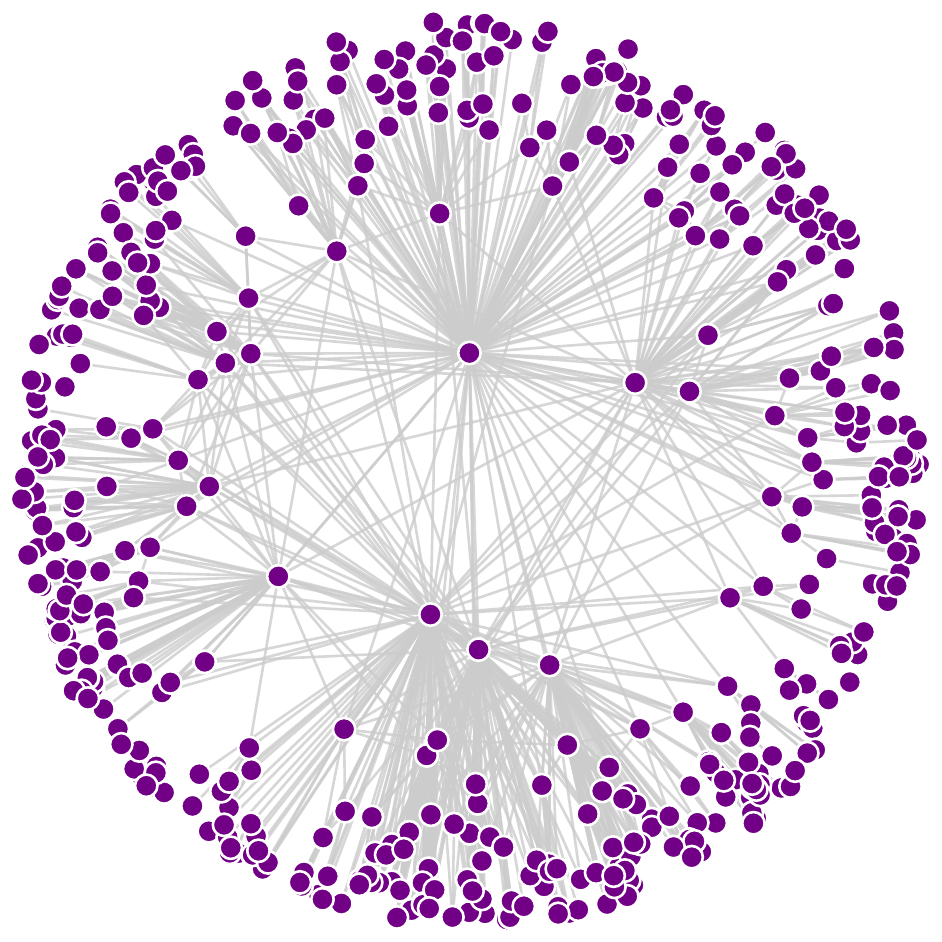}
	\caption{A hyperbolic random graph on 500 vertices with degree exponent $\tau=2.5$. Vertices are embedded based on their radial and angular coordinates.}
	\label{fig:hypergraph}
\end{figure}

The hyperbolic random graph samples $n$ vertices in a disk of radius $R=2\log(n/\nu)$, where the density of the radial coordinate $r$ of a vertex $p=(r,\phi)$ is
\begin{equation}
\rho(r)=\beta\frac{\sinh(\beta r)}{\cosh(\beta R)-1}
\end{equation}
with $\beta=(\tau-1)/2$. Here $\nu$ is a parameter that influences the average degree of the generated networks. The angle $\phi$ of $p$ is sampled uniformly from $[0,2\pi]$. Then, two vertices are connected if their hyperbolic distance is at most $R$. The hyperbolic distance of points $u=(r_u,\phi_u)$ and $v=(r_v,\phi_v)$ satisfies
\begin{align}
\cosh(\dd (u,v))=& \cosh(r_u)\cosh(r_v)-\sinh(r_u)\sinh(r_v)\cos(\Delta\theta).
\end{align}
 Two neighbors of a vertex are likely to be close to one another due to the geometric nature of the hyperbolic random graph. Therefore, the hyperbolic random graph contains many triangles~\cite{gugelmann2012}. Furthermore, the model generates scale-free networks with degree exponent $\tau$~\cite{krioukov2010} and small diameter~\cite{friedrich2015b}. Figure~\ref{fig:hypergraph} shows that vertices with small radial coordinates are often hubs, whereas vertices with larger radial coordinates usually have small degrees, which we explain in more detail in the Methods section. We use this relation between radial coordinate and degree to find the most likely triangle in the hyperbolic model in terms of degrees as well as radial coordinates. 

\subsection{Variational principle} 
The variational principle deals with 
the probability of creating a triangle between a vertex of degree $k$ and two other uniformly chosen vertices, which can be written as
\begin{equation}\label{eq:ptrpresent}
\Prob{\triangle_k}=\sum_{(d_1,d_2)}\Prob{\triangle\text{ on degrees }k,d_1,d_2}\Prob{d_1,d_2},
\end{equation}
where the sum is over all possible pairs of degrees $(d_1,d_2)$, and $\Prob{d_1,d_2}$ denotes the probability that two uniformly chosen vertices have degrees $d_1$ and $d_2$.
We then let the degrees $d_1$ and $d_2$ scale as $n^{\alpha_1}$ and $n^{\alpha_2}$ and find which degrees give the largest contribution to~\eqref{eq:ptrpresent}. Due to the power-law degree distribution, the probability that a vertex has degree proportional to $n^{\alpha}$ scales as $n^{-(\tau-1)\alpha}$. The maximal summand of~\eqref{eq:ptrpresent} can then be written as
\begin{equation}\label{eq:maxalphgen}
\max_{\alpha_1,\alpha_2}\Prob{\triangle\text{ on degrees }k,n^{\alpha_1},n^{\alpha_2}}n^{2+(\alpha_1+\alpha_2)(1-\tau)}.
\end{equation}
If the optimizer over $\alpha_1$ and $\alpha_2$ is unique, and attained by $\alpha_1^*$ and $\alpha_2^*$, then we can write the probability that a triangle is present between a vertex of degree $k$ and two randomly chosen vertices as
\begin{equation}\label{eq:ptropt}
\Prob{\triangle_k}\propto\Prob{\triangle\text{ on degrees }k,n^{\alpha_1^*},n^{\alpha_2^*}}n^{(\alpha_1^*+\alpha_2^*)(1-\tau)}.
\end{equation}
The local clustering coefficient $c(k)$ is defined as the expected number of triangles containing a uniformly chosen vertex of degree $k$ divided by $k\choose 2$. Therefore,
\begin{equation}\label{eq:ckalphmax}
{c(k)}\propto n^2k^{-2}\Prob{\triangle\text{ on degrees }k,n^{\alpha_1^*},n^{\alpha_2^*}}n^{(\alpha_1^*+\alpha_2^*)(1-\tau)}.
\end{equation}
Thus, if we know the probability that a triangle is present between vertices of degrees $k, n^{\alpha_1}$ and $n^{\alpha_2}$ for some random graph model, the variational principle is able to find the scaling of $c(k)$ in $k$ and the graph size $n$. 

Let us now explain why the variational principle~\eqref{eq:ckalphmax} applies to a fairly large class of random graphs. Suppose a model assigns to each vertex some parameters that determine the connection probabilities (radial and angular coordinates in case of the hyperbolic random graph). The variational principle can then be applied as long as the vertex degree can be expressed as some function of the vertex parameters, so that the probability of triangle formation between three vertices can be viewed as a function of the vertex degrees, and one can search for the optimal contribution to~\eqref{eq:maxalphgen}. 

\subsection{Local clustering} 
To compute $c(k)$ for the hyperbolic model, we calculate the probability that a triangle is present between vertices of degrees $k$, $n^{\alpha_1}$ and $n^{\alpha_2}$ using the variational principle. 

Vertices with small radial coordinates are often hubs, whereas vertices with larger radial coordinates usually have small degrees. We will use this relation between radial coordinate and degree to find the most likely triangle in the hyperbolic model in terms of degrees as well as radial coordinates. 
For a point $i$ with radial coordinate $r_i$, we define its type $t_i$ as
\begin{equation}
t_i=\me^{(R-r_i)/2}.
\end{equation}
Then, if $D_i$ denotes the degree of vertex $i$, by~\cite{stegehuis2017b}
\begin{equation}
t_i=\Theta(D_i).
\end{equation}
Furthermore, the $t_i$'s follow a power-law with exponent $\tau$~\cite{bode2015}, so that the degrees have a power-law distribution as well. The $t_i$'s can be interpreted as the weights in a hidden-variable model~\cite{bode2015}.

Because the degrees and the types of  vertices have the same scaling, we investigate the probability that two neighbors of a vertex of type $k$ connect. We compute the probability that a triangle is formed between a vertex of degree $k$, a vertex $i$ with $t_i\propto n^{\alpha_1}$ and a vertex $j$ with $t_j\propto n^{\alpha_2}$ with $\alpha_1\leq\alpha_2$. We can write this probability as
\begin{align}\label{eq:triangprob}
&\Prob{\triangle \text{ on types }k,n^{\alpha_1},n^{\alpha_2}}=\Prob{k\leftrightarrow n^{\alpha_1}}\Prob{k\leftrightarrow n^{\alpha_2}} \Prob{n^{\alpha_1}\text{ and }n^{\alpha_2}\text{ neighbors connect}}.
\end{align}
The probability that two vertices with types $t_i$ and $t_j$ connect satisfies by~\cite{bode2015}
\begin{equation}\label{eq:thetadif}
\Prob{i\leftrightarrow j\mid t_i,t_j} \propto \min\left({2\nu t_it_j}/{(\pi n)},1\right).
\end{equation}
Therefore, the probability that a vertex of type $k$ connects with a randomly chosen vertex of type $n^{\alpha_1}$ can be approximated by
\begin{equation}\label{eq:pkna}
\Prob{k \leftrightarrow   n^{\alpha_1}}\propto\min(k n^{\alpha_1-1},1).
\end{equation}


\begin{figure*}[tb]
	\centering
	\begin{subfigure}[t]{0.45\textwidth}
		\centering
		\includegraphics[width=0.7\textwidth]{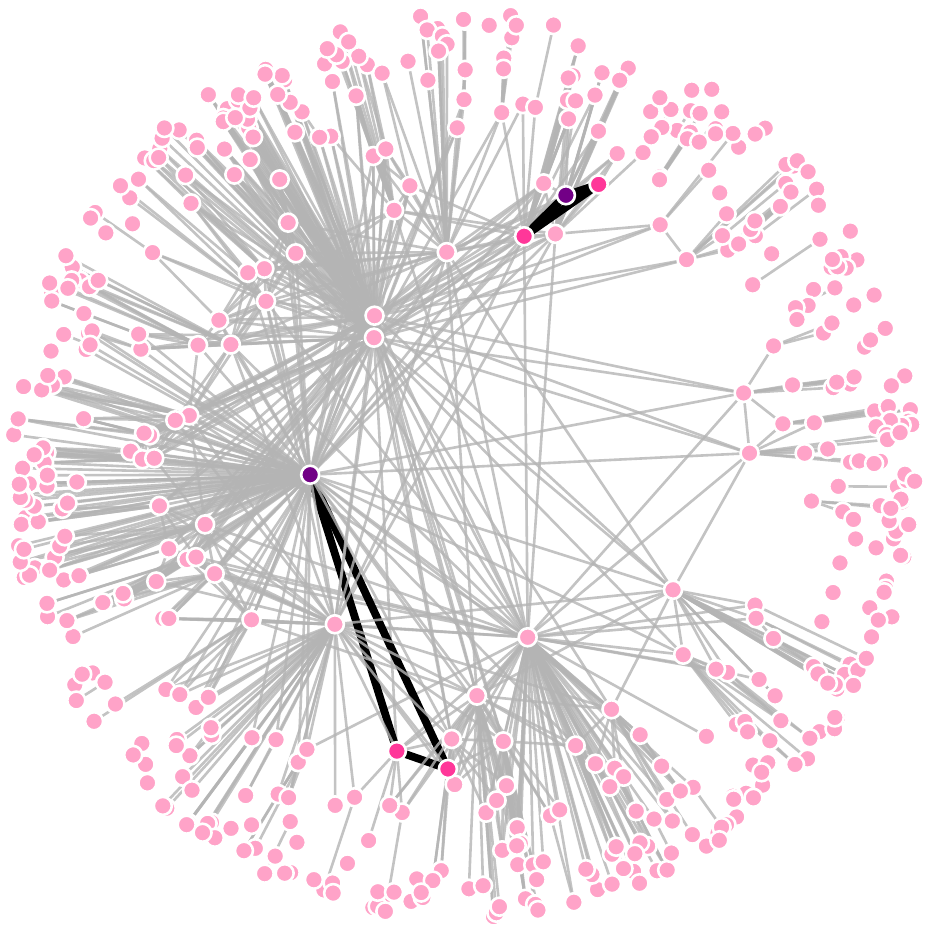}
		\caption{$\tau<5/2$: For $k\gg\sqrt{n}$ the two other vertices have degree proportional to $n/k$, whereas for $k\ll \sqrt{n}$ the other two vertices have degree proportional to $k$.}
		\label{fig:hyp22}
	\end{subfigure}
	\hspace{0.4cm}
	\begin{subfigure}[t]{0.45\textwidth}
		\centering
		\includegraphics[width=0.7\textwidth]{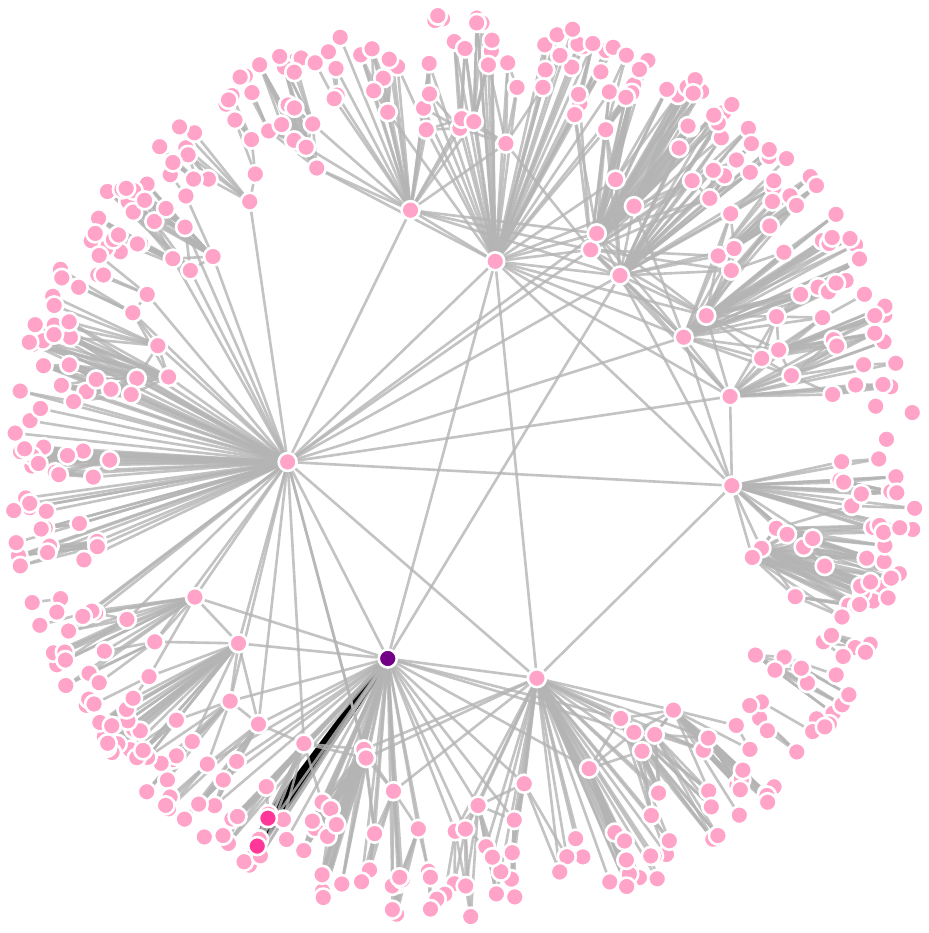}
		\caption{$\tau>5/2$: The other two vertices have constant degree across the entire range of $k$.}
		\label{fig:hyp28}
	\end{subfigure}
	\caption{Typical triangles containing a vertex of degree $k$ (dark red) in hyperbolic random graphs. A vertex of degree $n^{\alpha}$ has radial coordinate close to $R-\alpha$, so that the optimal triangle degrees can be translated back to their radial coordinates  in the disk. }
	\label{fig:hyptyp}
\end{figure*}

We now compute the order of magnitude of the third term in~\eqref{eq:triangprob}, which is more involved than the first two terms. Two neighbors of a vertex are likely to be close to one another, which increases the probability that they connect. 
Two vertices with types $t_i$ and $t_j$ and angular coordinates $\phi_i$ and $\phi_j$ connect if the relative angle between $\phi_i$ and $\phi_j$, $\Delta\theta$, satisfies~\cite{bode2015}
\begin{equation}\label{eq:relangle}
\Delta\theta \leq \Theta\left({2\nu t_it_j}/{n}\right).
\end{equation}
W.l.o.g., let the angular coordinate of the vertex with degree $k$ be 0. For $i$ and $j$ to be connected to a vertex with $\phi=0$, by~\eqref{eq:relangle} $\phi_i$ and $\phi_j$ must satisfy
\begin{equation}\label{eq:intervalsij}
\begin{aligned}[b]
&-\Theta(\min(kn^{\alpha_1-1},1))\leq \phi_i\leq \Theta(\min(kn^{\alpha_1-1},1)),\\
& -\Theta(\min(kn^{\alpha_2-1},1))\leq \phi_j\leq \Theta(\min(kn^{\alpha_2-1},1)).
\end{aligned}
\end{equation}
Because the angular coordinates in the hyperbolic random graph are uniformly distributed, $\phi_i$ and $\phi_j$ are uniformly distributed in these ranges. By~\eqref{eq:relangle}, vertices $i$ and $j$ are connected if their relative angle is at most
\begin{equation}\label{eq:ijtheta}
2\nu n^{\alpha_1+\alpha_2-1}.
\end{equation}
Thus, the probability that $i$ and $j$ connect is the probability that two randomly chosen points in the intervals~\eqref{eq:intervalsij} differ in their angles by at most~\eqref{eq:ijtheta}. 
Assume that $\alpha_2\geq\alpha_1$. Then, the probability that $i$ and $j$ are connected is proportional to
\begin{align}
\Prob{n^{\alpha_1}\text{ and }n^{\alpha_2}\text{ neighbor connect}} & \propto \min\bigg(\frac{n^{\alpha_1+\alpha_2-1}}{\min(kn^{\alpha_2-1},1)},1\bigg)\nonumber\\
& =\min(n^{\alpha_1}\max(n^{\alpha_2-1},k^{-1}),1).
\end{align}
Thus,~\eqref{eq:maxalphgen} reduces to
\begin{align}
\max_{\alpha_1,\alpha_2} &  \  n^{(\alpha_1+\alpha_2)(1-\tau)}\min(kn^{\alpha_1-1},1)\min(kn^{\alpha_2-1},1) \min(n^{\alpha_1}\max(n^{\alpha_2-1},k^{-1}),1).
\end{align}
Because of the $\min(kn^{\alpha_2-1},1)$ term, it is never optimal to let the $\max$ term be attained by $n^{\alpha_2-1}$. Thus, the equation reduces further to
\begin{align}\label{eq:hypermax}
\max_{\alpha_1,\alpha_2} & \  n^{(\alpha_1+\alpha_2)(1-\tau)}\min(kn^{\alpha_1-1},1)\min(kn^{\alpha_2-1},1) \min(n^{\alpha_1}k^{-1},1).
\end{align}
The maximizers over $\alpha_1\leq \alpha_2$ are given by
\begin{equation}\label{eq:ckhypermax}
(n^{\alpha_1},n^{\alpha_2})\propto \begin{cases}
(n^0,n^0), & \tau>\tfrac{5}{2},\\
(k,k) & \tau<\tfrac{5}{2}, k\ll \sqrt{n},\\
(n/k,n/k) & \tau<\tfrac{5}{2}, k\gg \sqrt{n}.
\end{cases}
\end{equation}
Combining this with~\eqref{eq:ckalphmax} shows that
\begin{equation}\label{eq:ckhyp}
c(k)\propto
\begin{cases}
k^{-1}& \tau>\tfrac{5}{2},\\
k^{4-2\tau} & \tau<\tfrac{5}{2}, k\ll\sqrt{n},\\
k^{2\tau-6}n^{5-2\tau} & \tau<\tfrac{5}{2},k\gg\sqrt{n}.
\end{cases}
\end{equation}

\begin{figure*}[tb]
	\centering
	\begin{subfigure}[t]{0.31\textwidth}
		\includegraphics[width=\textwidth]{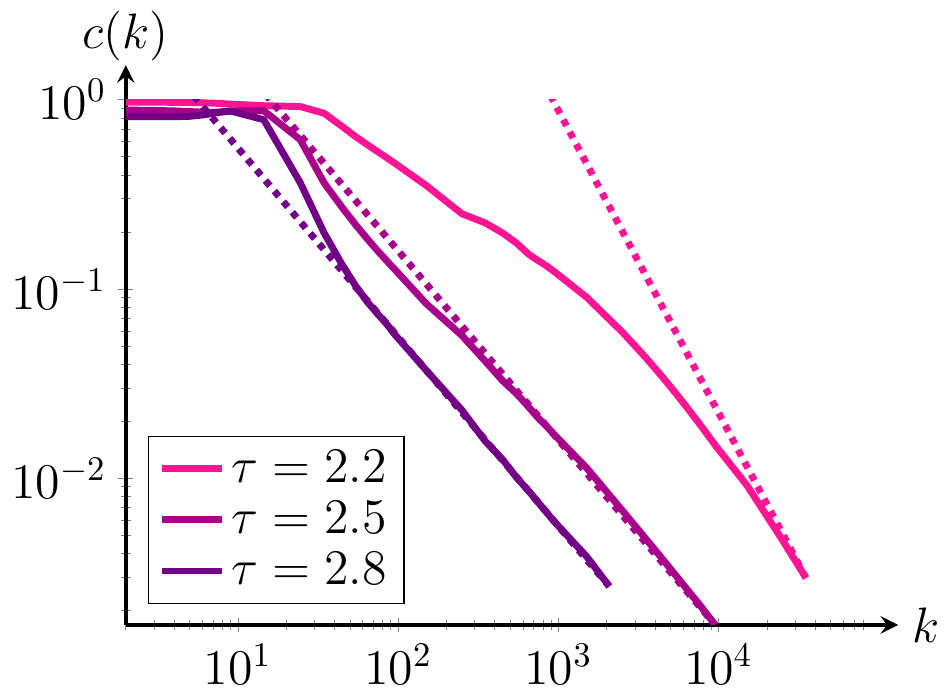}
		\caption{Hyperbolic random graph with $\nu=1$.}
		\label{fig:cksim}
	\end{subfigure}
	\hspace{0.1cm}
	\begin{subfigure}[t]{0.32\linewidth}
		\centering
		\includegraphics[width=0.9\textwidth]{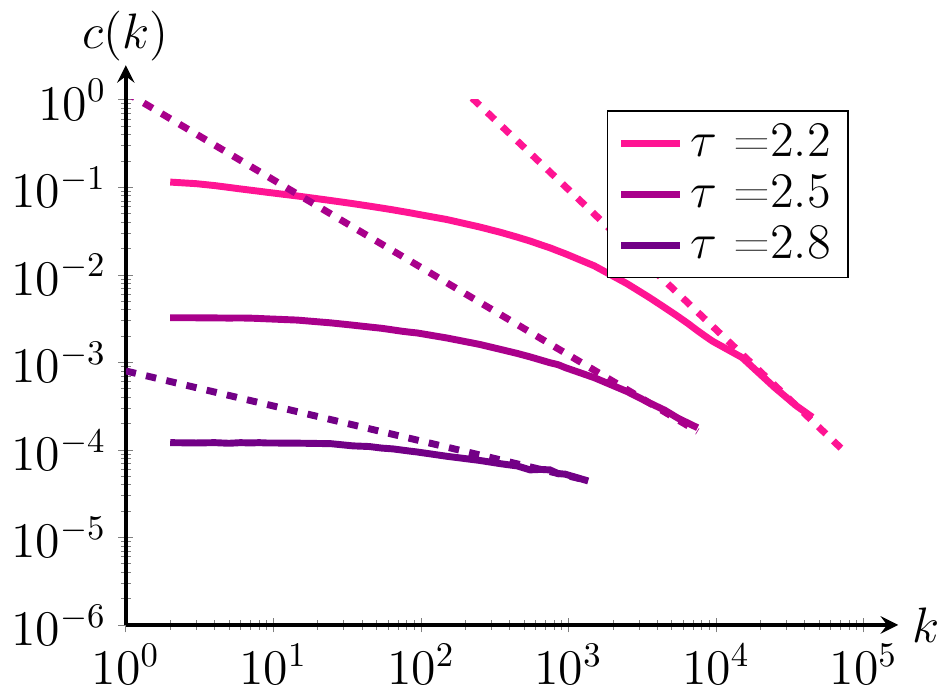}
		\caption{Hidden-variable model}
		\label{fig:hidck}
	\end{subfigure}
	\hspace{0.1cm}
	\begin{subfigure}[t]{0.32\linewidth}
		\centering
		\includegraphics[width=0.9\textwidth]{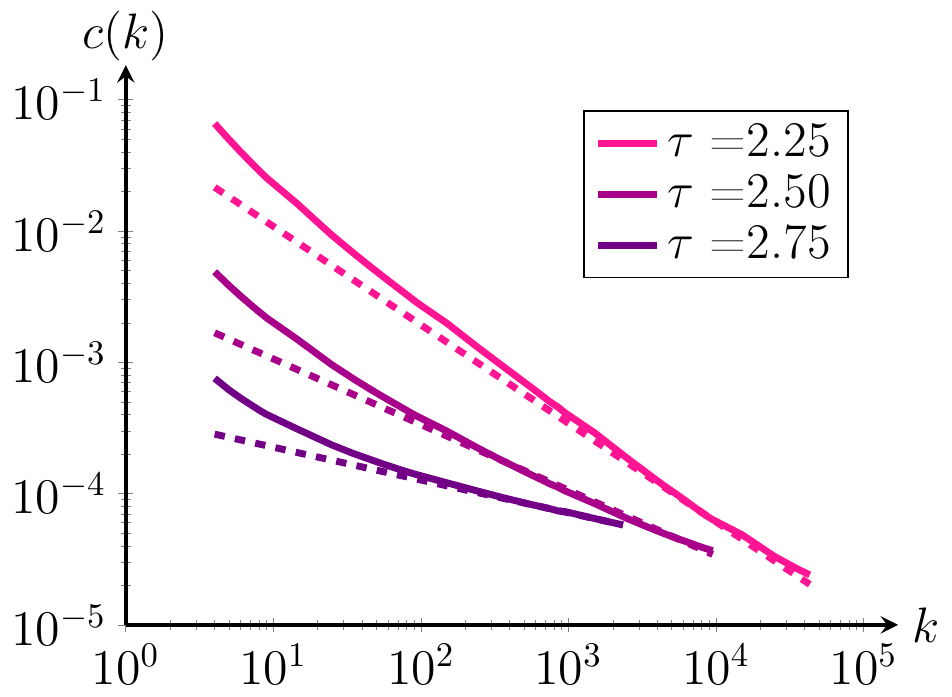}
		\caption{Preferential attachment model with $m=4$}
		\label{fig:PAck}
	\end{subfigure}
	\caption{Simulations of $c(k)$ for three different models with $n=10^6$. The solid lines correspond to averages over $10^4$ network realizations and the dashed lines indicate the asymptotic slopes of~\eqref{eq:ckhyp},~\eqref{eq:ckknownmain} and~\eqref{eq:ckpa}.}
	\label{fig:cksims}
\end{figure*}

This result is more detailed than the result in~\cite{krioukov2010}, where the scaling $c(k)\sim k^{-1}$ was predicted for fixed $k$. We find that this scaling only holds for the larger values of $\tau$, while for $\tau<5/2$ the decay of the curve is significantly different, which was simultaneously found in~\cite{hoorn2018}. Note that for $\tau> 5/2$, the $c(k)$ curve does not depend on $n$. For $\tau<5/2$ the dependence on $n$ is only present for large values of $k$. Interestingly, the exponent $\tau=5/2$ is also the point where the maximal contribution to a bidirectional shortest path in the hyperbolic random graph changes from high-degree to lower-degree vertices~\cite{blasius2018}. Also, the optimal triangle structures contain higher vertex degrees for $\tau<2.5$ than for $\tau>2.5$ (see Fig.~\ref{fig:hyptyp}).
Figure~\ref{fig:cksim} confirms the asymptotic slopes~\eqref{eq:ckhyp} with extensive simulations.

\section{Self-averaging behavior}\label{ssf}
We say that  $c(k)$ is self-averaging when $\Var{c(k)}/\Exp{c(k)}^2\to 0$ as $n\to\infty$, so that the sample-to-sample fluctuations of $c(k)$ vanish in the large-network limit. When $c(k)$ fails to be self-averaging, the fluctuations persist even in the large-network limit, so that the average of $c(k)$ over many network realizations cannot be viewed as a reliable descriptor of local clustering. 
We will now show how to apply the variational principle~\eqref{eq:ckalphmax} to constrained subgraphs larger than triangles, which leads to a complete characterization of $\Var{c(k)}/\Exp{c(k)}^2$ in the large-network limit. The variational principle can hence determine for any value of $k$ whether $c(k)$ is self-averaging or not. 
In this way we are able to show that for the hyperbolic random graph, \emph{$c(k)$ is self-averaging for all values of $\tau\in(2,3)$ and all $k$}. This implies that for large enough $n$, one sample of the hyperbolic random graph is sufficient to obtain the characteristic behavior of $c(k)$. 

\subsection{Extended variational principle}
To show that $c(k)$ is self-averaging, we first study $\Exp{c(k)}$. 
In the variational principle~\eqref{eq:maxalphgen}, we obtained the typical number of triangles where one vertex has degree $k$ by putting the hard constraint $\alpha_1,\alpha_2\leq {1}/(\tau-1)$ on the degrees of the other two vertices in the triangle. If we relax this constraint, we can compute $\Exp{c(k)}$. This quantity can be interpreted as the value of $c(k)$ obtained after simulating many hyperbolic random graphs, and taking the average value of $c(k)$ over all these hidden-variable models.
We see from~\eqref{eq:ckhypermax} that the largest contribution to $c(k)$ is from vertices with degrees strictly smaller than $n^{1/(\tau-1)}$. Thus, removing the constraint on the maximal degree does not influence the major contribution, so that, similarly to~\eqref{eq:ckhyp},
\begin{equation}\label{eq:expckhyp}
\Exp{c(k)}\propto
\begin{cases}
k^{-1}& \tau>\tfrac{5}{2},\\
k^{4-2\tau} & \tau<\tfrac{5}{2}, k\ll\sqrt{n},\\
k^{2\tau-6}n^{5-2\tau} & \tau<\tfrac{5}{2},k\gg\sqrt{n}.
\end{cases}
\end{equation}

We now compute the variance of $c(k)$. 
Note that
\begin{equation}\label{eq:varck}
\Var{c(k)}=\frac{1}{k^4N_k^2}\Var{\triangle_k},
\end{equation}
where $\triangle_k$ denotes the total number of triangles attached to a vertex of degree $k$.
The variance of $\triangle_k$ can be computed as
\begin{align}\label{eq:varconstrained}
\Var{\triangle_k}& = \sum_{i,j:d_i,d_j=k}\sideset{}{'}\sum_{u,v\in[n]}\sideset{}{'}\sum_{w,z\in[n]}\Prob{\triangle_{i,u,v}\triangle_{j,w,z}} -\Prob{\triangle_{i,u,v}}\Prob{\triangle_{j,w,z}}.
\end{align}
When $i,u,v$ and $j,w,z$ do not overlap, their weights are independent, so that the events that $i,u$ and $v$ form a triangle and that $j,w$ and $z$ form a triangle are independent. Thus, when $i,j,u,v,w,z$ are distinct indices, $\Prob{\triangle_{i,u,v}\triangle_{j,w,z}}=\Prob{\triangle_{i,u,v}}\Prob{\triangle_{j,w,z}}$, so that the contribution from 6 distinct indices to~\eqref{eq:varconstrained} is zero. Similarly, when $i=j$, the weight of $i$ is $k(1+\op(1))$. Therefore, $\Prob{\triangle_{i,u,v}\triangle_{i,w,z}}=\Prob{\triangle_{i,u,v}}\Prob{\triangle_{j,w,z}}(1+\op(1))$ as long as $u,v,w,z$ are distinct. Thus, the contribution to the variance of $\triangle_k$ from $i=j$ can be bounded as $o(\Exp{c(k)}^2)$.

On the other hand, when $u=w$ for example, the first term in~\eqref{eq:varconstrained} denotes the probability that a bow-tie is present with $u$ as middle vertex. Furthermore, since the degrees are i.i.d., for any $i\neq u\neq v$, such that $d_i=k$,
\begin{equation}
\Prob{\triangle_{i,u,v}}= \frac{\Exp{\triangle_k}}{6{n\choose 3}}.
\end{equation}
This results in
\begin{align}\label{eq:varconst}
\Var{\triangle_k} &=4 \Exp{\bowa}+4 \Exp{\bowb}+\Exp{\bowc}+2\Exp{\diamonda }+4\Exp{\diamondb} +8\Exp{\diamondc} +4\Exp{\diamondd} \nonumber\\
& \quad +2\Exp{\triangle_k}+4\Exp{\triangd}+\Exp{\triangle_k}^2O(n^{-1})
\end{align}
where $\scalebox{0.8}{\bowa}$ denotes a bow-tie where the white vertices are constrained to have degree $k$.
The combinatorial factor 4 arises in the first term because there are 4 ways to construct a bow-tie where two constrained vertices have degree $k$ by letting two triangles containing a degree $k$ vertex overlap. The other combinatorial factors arise similarly.

We write the first expectation as
\begin{equation}
\Exp{\bowa}=n^3N_k^2\Prob{\bowa},
\end{equation}
where $\prob\big(\scalebox{0.8}{\bowa}\big)$ denotes the probability that two randomly chosen vertices of degree $k$ form the constrained bow-tie together with three randomly chosen other vertices, and $N_k$ denotes the number of degree-$k$ vertices. We can compute this probability with a constrained variational principle. By symmetry of the bow-tie subgraph, the optimal degree range of the bottom right vertex and the upper right vertex is the same. Let the degree of the middle vertex scale as $n^{\alpha_1}$, and the degrees of the other two vertices as $n^{\alpha_2}$. Then, we write the constrained variational principle, similarly to~\eqref{eq:hypermax}, as
\begin{align}\label{eq:bowtiek2}
& n^3N_k^2n^{(2\alpha_1+\alpha_2)(1-\tau)}\min(kn^{\alpha_1-1},1)^2\min(kn^{\alpha_2-1},1)^2 \min(n^{\alpha_1}\max(n^{\alpha_2-1},k^{-1}),1)^2.
\end{align}
Optimizing this over $\alpha_1$ and $\alpha_2$ yields that for $k\ll\sqrt{n}$ the number of $\scalebox{0.8}{\bowa}$ subgraphs is dominated by the type displayed in Fig.~\ref{fig:bowk1hyp}, where $n^{\alpha_1}\propto k$ and $n^{\alpha_2}\propto n/k$. Computing this contribution results in
\begin{equation}\label{eq:bowaexp}
\Exp{\bowa}\propto n^3N_k^2k^{2(1-\tau)}\left(\frac{n}{k}\right)^{1-\tau}\Big(\frac{k^2}{n}\Big)^2= N_k^2n^{2-\tau}k^{5-\tau}.
\end{equation}
Thus, using~\eqref{eq:varck} shows that the contribution to the variance is $n^{2-\tau}k^{1-\tau}$, as shown in Fig.~\ref{fig:bowk1hyp}. We obtain using~\eqref{eq:expckhyp} that for $k\ll \sqrt{n}$,
\begin{equation}
\frac{n^{2-\tau}k^{1-\tau}}{\Exp{c(k)}^2}\propto
\begin{cases}
n^{2-\tau}k^{3-\tau} \ll n^{(7-3\tau)/2}& \tau>\tfrac{5}{2},\\
n^{2-\tau}k^{3\tau-7} \ll \max(n^{(\tau-3)/2},n^{2-\tau})& \tau<\tfrac{5}{2},
\end{cases}
\end{equation}
which tends to zero as $n\to\infty$. Thus, the contribution to the variance from the $\scalebox{0.8}{\bowa}$ subgraphs tends to zero in the large network limit for $k\ll \sqrt{n}$.

The optimizer of~\eqref{eq:bowtiek2} for $k\gg\sqrt{n}$ is for $n^{\alpha_1}\propto n/k$ and $n^{\alpha_2}\propto n/k$. Thus, similarly to~\eqref{eq:bowaexp}
\begin{equation}\label{eq:bowaexpl}
\Exp{\bowa}\propto n^3N_k^2\left(\frac{n}{k}\right)^{3(1-\tau)}\left(\frac{n}{k^2}\right)^2= N_k^2n^{8-3\tau}k^{3\tau-7}.
\end{equation}
The contribution to the variance then is $n^{8-3\tau}k^{3\tau-11}$, as Fig.~\ref{fig:bowk1hyps} shows. Thus, for $k\gg\sqrt{n}$,
\begin{equation}
\frac{n^{8-3\tau}k^{3\tau-11}}{\Exp{c(k)}^2} \propto \begin{cases}
n^{8-3\tau}k^{3(\tau-3)} \ll n^{(7-3\tau)/2}& \tau>\tfrac{5}{2},\\
n^{\tau-2}k^{1-\tau} \ll n^{(\tau-3)/2}& \tau<\tfrac{5}{2},
\end{cases}
\end{equation}
which tends to zero as $n\to\infty$, showing that indeed the contribution from the $\scalebox{0.8}{\bowa}$ subgraph to the variance is small when $k\gg\sqrt{n}$. 

The contributions of other types of merged triangles to the variance of $c(k)$ can be computed similarly, and are shown in Figs~\ref{fig:bowtiehyp} and~\ref{fig:bowtiekhyps}.
 These contributions are all smaller than $\Exp{c(k)}^2$ (see~\eqref{eq:expckhyp}), so that $c(k)$ is self-averaging over its entire spectrum. 

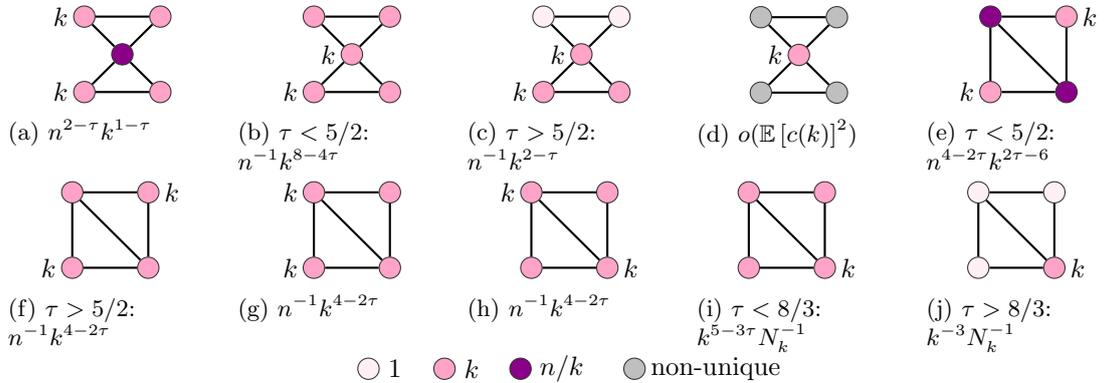
\begin{figure}[htb]
	\captionsetup{justification=raggedright,singlelinecheck=false}
	\tikzstyle{every node}=[circle,fill=black!25,minimum size=8pt,inner sep=0pt,draw=black!80]
	\centering
	\begin{subfigure}[t]{0.18\linewidth}
		\centering
		\begin{tikzpicture}
		\centering
		\tikzstyle{edge} = [draw,thick,-]
		\node[S1,label=left:{$k$}] (a) at (0,0) {};
		\node[S1] (b) at (1,0) {};
		\node[S1,label=left:{$k$}] (c) at (0,1) {};
		\node[S1] (d) at (1,1) {};
		\node[S2m] (e) at (0.5,0.5) {};
		\draw[edge] (a)--(b);
		\draw[edge] (e)--(a);
		\draw[edge] (b)--(e);
		\draw[edge] (e)--(d);
		\draw[edge] (c)--(e);
		\draw[edge] (c)--(d);
		\end{tikzpicture}
		\caption{$n^{2-\tau}k^{1-\tau}$}
		\label{fig:bowk1hyp}
	\end{subfigure}	
	\hspace{0.1cm}
	\begin{subfigure}[t]{0.18\linewidth}
		\centering
		\begin{tikzpicture}
		\tikzstyle{edge} = [draw,thick,-]
		\node[S1,label=left:{$k$}] (a) at (0,0) {};
		\node[S1] (b) at (1,0) {};
		\node[S1] (c) at (0,1) {};
		\node[S1] (d) at (1,1) {};
		\node[S1,label=left:{$k$}] (e) at (0.5,0.5) {};
		\draw[edge] (a)--(b);
		\draw[edge] (e)--(a);
		\draw[edge] (b)--(e);
		\draw[edge] (e)--(d);
		\draw[edge] (c)--(e);
		\draw[edge] (c)--(d);
		\end{tikzpicture}
		\caption{$\tau<5/2$: $n^{-1}k^{8-4\tau}$}
		\label{fig:bowk2hypsmall}
	\end{subfigure}	
	\hspace{0.1cm}
	\begin{subfigure}[t]{0.18\linewidth}
		\centering
		\begin{tikzpicture}
		\tikzstyle{edge} = [draw,thick,-]
		\node[S1,label=left:{$k$}] (a) at (0,0) {};
		\node[S1] (b) at (1,0) {};
		\node[n1] (c) at (0,1) {};
		\node[n1] (d) at (1,1) {};
		\node[S1,label=left:{$k$}] (e) at (0.5,0.5) {};
		\draw[edge] (a)--(b);/
		\draw[edge] (e)--(a);
		\draw[edge] (b)--(e);
		\draw[edge] (e)--(d);
		\draw[edge] (c)--(e);
		\draw[edge] (c)--(d);
		\end{tikzpicture}
		\caption{$\tau>5/2$: $n^{-1}k^{2-\tau}$}
		\label{fig:bowk2hyplarge}
	\end{subfigure}	
	\hspace{0.1cm}
	\begin{subfigure}[t]{0.18\linewidth}
		\centering
		\begin{tikzpicture}
		\tikzstyle{edge} = [draw,thick,-]
		\node[] (a) at (0,0) {};
		\node[] (b) at (1,0) {};
		\node[] (c) at (0,1) {};
		\node[] (d) at (1,1) {};
		\node[S1,label=left:{$k$}] (e) at (0.5,0.5) {};
		\draw[edge] (a)--(b);
		\draw[edge] (e)--(a);
		\draw[edge] (b)--(e);
		\draw[edge] (e)--(d);
		\draw[edge] (c)--(e);
		\draw[edge] (c)--(d);
		\end{tikzpicture}
		\caption{$o(\Exp{c(k)}^2)$}
		\label{fig:bowk3hyp}
	\end{subfigure}	
\hspace{0.1cm}
	\begin{subfigure}[t]{0.18\linewidth}
		\centering
		\begin{tikzpicture}
		\tikzstyle{edge} = [draw,thick,-]
		\node[S1,label=left:{$k$}] (a) at (0,0) {};
		\node[S2m] (b) at (1,0) {};
		\node[S2m] (c) at (0,1) {};
		\node[S1,label=right:{$k$}] (d) at (1,1) {};
		\draw[edge] (a)--(b);
		\draw[edge] (c)--(b);
		\draw[edge] (d)--(b);
		\draw[edge] (a)--(c);
		\draw[edge] (c)--(d);
		\end{tikzpicture}	
		\caption{$\tau<5/2$: $n^{4-2\tau}k^{2\tau-6}$}
		\label{fig:diamondk1hypsmall}
	\end{subfigure}

	\begin{subfigure}[t]{0.18\linewidth}
		\centering
		\begin{tikzpicture}
		\tikzstyle{edge} = [draw,thick,-]
		\node[S1,label=left:{$k$}] (a) at (0,0) {};
		\node[S1] (b) at (1,0) {};
		\node[S1] (c) at (0,1) {};
		\node[S1,label=right:{$k$}] (d) at (1,1) {};
		\draw[edge] (a)--(b);
		\draw[edge] (c)--(b);
		\draw[edge] (d)--(b);
		\draw[edge] (a)--(c);
		\draw[edge] (c)--(d);
		\end{tikzpicture}	
		\caption{$\tau>5/2$: $n^{-1}k^{4-2\tau}$}
		\label{fig:diamondk1hyp}
	\end{subfigure}
	\hspace{0.1cm}
	\begin{subfigure}[t]{0.18\linewidth}
		\centering
		\begin{tikzpicture}
		\tikzstyle{edge} = [draw,thick,-]
		\node[S1,label=left:{$k$}] (a) at (0,0) {};
		\node[S1] (b) at (1,0) {};
		\node[S1,label=left:{$k$}] (c) at (0,1) {};
		\node[S1] at (1,1) {};
		\draw[edge] (a)--(b);
		\draw[edge] (c)--(b);
		\draw[edge] (d)--(b);
		\draw[edge] (a)--(c);
		\draw[edge] (c)--(d);
		\end{tikzpicture}	
		\caption{$n^{-1}k^{4-2\tau}$}
		\label{fig:diamondk2hyp}
	\end{subfigure}
	\hspace{0.1cm}
	\begin{subfigure}[t]{0.18\linewidth}
		\centering
		\begin{tikzpicture}
		\tikzstyle{edge} = [draw,thick,-]
		\node[S1] (a) at (0,0) {};
		\node[S1,label=right:{$k$}] (b) at (1,0) {};
		\node[S1,label=left:{$k$}] (c) at (0,1) {};
		\node[S1] at (1,1) {};
		\draw[edge] (a)--(b);
		\draw[edge] (c)--(b);
		\draw[edge] (d)--(b);
		\draw[edge] (a)--(c);
		\draw[edge] (c)--(d);
		\end{tikzpicture}	
		\caption{$n^{-1}k^{4-2\tau}$}
		\label{fig:diamondk3hyp }
	\end{subfigure}
\hspace{0.1cm}	
	\begin{subfigure}[t]{0.18\linewidth}
		\centering
		\begin{tikzpicture}
		\tikzstyle{edge} = [draw,thick,-]
		\node[S1] (a) at (0,0) {};
		\node[S1,label=right:{$k$}](b) at (1,0) {};
		\node[S1] (c) at (0,1) {};
		\node[S1] (d) at (1,1) {};
		\draw[edge] (a)--(b);
		\draw[edge] (c)--(b);
		\draw[edge] (d)--(b);
		\draw[edge] (a)--(c);
		\draw[edge] (c)--(d);
		\end{tikzpicture}	
		\caption{$\tau<8/3$: $k^{5-3\tau}N_k^{-1}$}
		\label{fig:diamondk4hypsmall}
	\end{subfigure}
	\hspace{0.1cm}
	\begin{subfigure}[t]{0.18\linewidth}
		\centering
		\begin{tikzpicture}
		\tikzstyle{edge} = [draw,thick,-]
		\node[n1] (a) at (0,0) {};
		\node[S1,label=right:{$k$}](b) at (1,0) {};
		\node[n1] (c) at (0,1) {};
		\node[n1] (d) at (1,1) {};
		\draw[edge] (a)--(b);
		\draw[edge] (c)--(b);
		\draw[edge] (d)--(b);
		\draw[edge] (a)--(c);
		\draw[edge] (c)--(d);
		\end{tikzpicture}	
		\caption{$\tau>8/3$: $k^{-3}N_k^{-1}$ }
		\label{fig:diamondk4hyp}
	\end{subfigure}
	
	\vspace{-0.7cm}
	\begin{subfigure}{\linewidth}
		\centering
		\begin{tikzpicture}
		\node[S2m,label={[label distance=0.05cm]0:$n/k$}] (a) at (2,0) {};
		\node[S1,label={[label distance=0.05cm]0:$k$}] (c) at (1,0) {};
		\node[n1,label={[label distance=0.05cm]0:$1$}] (c) at (0,0) {};
		\node[label={[label distance=0.05cm]0:non-unique}] (d) at (3.5,0) {};
		\end{tikzpicture}
	\end{subfigure}
	\vspace{-0.8cm}
	
	\caption{Contribution to the variance of $c(k)$ in the hyperbolic model from merging two triangles where one vertex has degree $k\ll \sqrt{n}$. The vertex color indicates the optimal vertex degree.}
	\label{fig:bowtiehyp}
\end{figure}
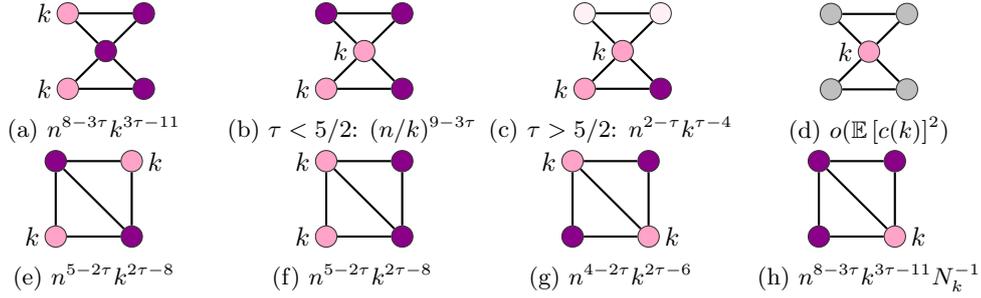
\begin{figure}[htb]
	\tikzstyle{every node}=[circle,fill=black!25,minimum size=8pt,inner sep=0pt,draw=black!80]
	\captionsetup[subcaption]{justification=centering,singlelinecheck=false}
	\centering
	\begin{subfigure}[t]{0.22\linewidth}
		\centering
		\begin{tikzpicture}
		\centering
		\tikzstyle{edge} = [draw,thick,-]
		\node[S1,label=left:{$k$}] (a) at (0,0) {};
		\node[S2m] (b) at (1,0) {};
		\node[S1,label=left:{$k$}] (c) at (0,1) {};
		\node[S2m] (d) at (1,1) {};
		\node[S2m] (e) at (0.5,0.5) {};
		\draw[edge] (a)--(b);
		\draw[edge] (e)--(a);
		\draw[edge] (b)--(e);
		\draw[edge] (e)--(d);
		\draw[edge] (c)--(e);
		\draw[edge] (c)--(d);
		\end{tikzpicture}
		\caption{$n^{8-3\tau}k^{3\tau-11}$}
		\label{fig:bowk1hyps}
	\end{subfigure}	
	\begin{subfigure}[t]{0.22\linewidth}
		\centering
		\begin{tikzpicture}
		\tikzstyle{edge} = [draw,thick,-]
		\node[S1,label=left:{$k$}] (a) at (0,0) {};
		\node[S2m] (b) at (1,0) {};
		\node[S2m] at (0,1) {};
		\node[S2m] at (1,1) {};
		\node[S1,label=left:{$k$}] (e) at (0.5,0.5) {};
		\draw[edge] (a)--(b);
		\draw[edge] (e)--(a);
		\draw[edge] (b)--(e);
		\draw[edge] (e)--(d);
		\draw[edge] (c)--(e);
		\draw[edge] (c)--(d);
		\end{tikzpicture}
		\caption{$\tau<5/2$: $(n/k)^{9-3\tau}$}
		\label{fig:bowk2hypsmalls}
	\end{subfigure}	
	\begin{subfigure}[t]{0.22\linewidth}
		\centering
		\begin{tikzpicture}
		\tikzstyle{edge} = [draw,thick,-]
		\node[S1,label=left:{$k$}] (a) at (0,0) {};
		\node[S2m] (b) at (1,0) {};
		\node[n1] at (0,1) {};
		\node[n1] at (1,1) {};
		\node[S1,label=left:{$k$}] (e) at (0.5,0.5) {};
		\draw[edge] (a)--(b);
		\draw[edge] (e)--(a);
		\draw[edge] (b)--(e);
		\draw[edge] (e)--(d);
		\draw[edge] (c)--(e);
		\draw[edge] (c)--(d);
		\end{tikzpicture}
		\caption{$\tau>5/2$: $n^{2-\tau}k^{\tau-4}$}
		\label{fig:bowk2hyps}
	\end{subfigure}	
	\begin{subfigure}[t]{0.22\linewidth}
		\centering
		\begin{tikzpicture}
		\tikzstyle{edge} = [draw,thick,-]
		\node[] (a) at (0,0) {};
		\node[] (b) at (1,0) {};
		\node[] (c) at (0,1) {};
		\node[] (d) at (1,1) {};
		\node[S1,label=left:{$k$}] (e) at (0.5,0.5) {};
		\draw[edge] (a)--(b);
		\draw[edge] (e)--(a);
		\draw[edge] (b)--(e);
		\draw[edge] (e)--(d);
		\draw[edge] (c)--(e);
		\draw[edge] (c)--(d);
		\end{tikzpicture}
		\caption{$o(\Exp{c(k)}^2)$}
		\label{fig:bowk3hyps}
	\end{subfigure}	
	
	\begin{subfigure}[t]{0.22\linewidth}
		\centering
		\begin{tikzpicture}
		\tikzstyle{edge} = [draw,thick,-]
		\node[S1,label=left:{$k$}] (a) at (0,0) {};
		\node[S2m] (b) at (1,0) {};
		\node[S2m] (c) at (0,1) {};
		\node[S1,label=right:{$k$}] (d) at (1,1) {};
		\draw[edge] (a)--(b);
		\draw[edge] (c)--(b);
		\draw[edge] (d)--(b);
		\draw[edge] (a)--(c);
		\draw[edge] (c)--(d);
		\end{tikzpicture}	
		\caption{$n^{5-2\tau}k^{2\tau-8}$}
		\label{fig:diamondk1hyps}
	\end{subfigure}
	\begin{subfigure}[t]{0.22\linewidth}
		\centering
		\begin{tikzpicture}
		\tikzstyle{edge} = [draw,thick,-]
		\node[S1,label=left:{$k$}] (a) at (0,0) {};
		\node[S2m] (b) at (1,0) {};
		\node[S1,label=left:{$k$}] (c) at (0,1) {};
		\node[S2m] at (1,1) {};
		\draw[edge] (a)--(b);
		\draw[edge] (c)--(b);
		\draw[edge] (d)--(b);
		\draw[edge] (a)--(c);
		\draw[edge] (c)--(d);
		\end{tikzpicture}	
		\caption{$n^{5-2\tau}k^{2\tau-8}$}
		\label{fig:diamondk2hyps}
	\end{subfigure}
	\begin{subfigure}[t]{0.22\linewidth}
		\centering
		\begin{tikzpicture}
		\tikzstyle{edge} = [draw,thick,-]
		\node[S2m] (a) at (0,0) {};
		\node[S1,label=right:{$k$}] (b) at (1,0) {};
		\node[S1,label=left:{$k$}] (c) at (0,1) {};
		\node[S2m] at (1,1) {};
		\draw[edge] (a)--(b);
		\draw[edge] (c)--(b);
		\draw[edge] (d)--(b);
		\draw[edge] (a)--(c);
		\draw[edge] (c)--(d);
		\end{tikzpicture}	
		\caption{$n^{4-2\tau}k^{2\tau-6}$}
		\label{fig:diamondk3hyps}
	\end{subfigure}
	\begin{subfigure}[t]{0.22\linewidth}
		\centering
		\begin{tikzpicture}
		\tikzstyle{edge} = [draw,thick,-]
		\node[S2m] (a) at (0,0) {};
		\node[S1,label=right:{$k$}](b) at (1,0) {};
		\node[S2m] (c) at (0,1) {};
		\node[S2m] (d) at (1,1) {};
		\draw[edge] (a)--(b);
		\draw[edge] (c)--(b);
		\draw[edge] (d)--(b);
		\draw[edge] (a)--(c);
		\draw[edge] (c)--(d);
		\end{tikzpicture}	
		\caption{$n^{8-3\tau}k^{3\tau-11}N_k^{-1}$}
		\label{fig:diamondk4hyps}
	\end{subfigure}
	\caption{Contribution to the variance of $c(k)$ in the hyperbolic model from merging two triangles where one vertex has degree $k\gg \sqrt{n}$. The vertex color indicates the optimal vertex degree as in Fig.~\ref{fig:bowtiehyp}.}
	\label{fig:bowtiekhyps}
\end{figure}

\subsection{Global clustering}
\begin{figure}[htb]
	\centering
	\begin{subfigure}{0.25\textwidth}
		\includegraphics[width=\textwidth]{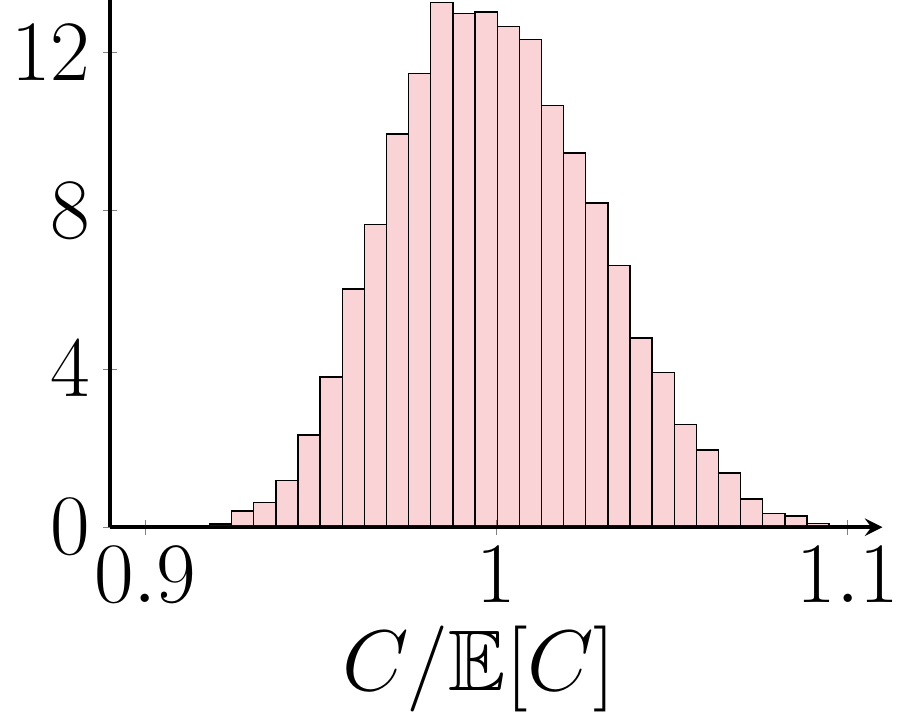}
		\caption{$n=10^4,\tau=2.2$}
		\label{fig:Chyp2210000}
	\end{subfigure}
	\begin{subfigure}{0.25\textwidth}
		\includegraphics[width=\textwidth]{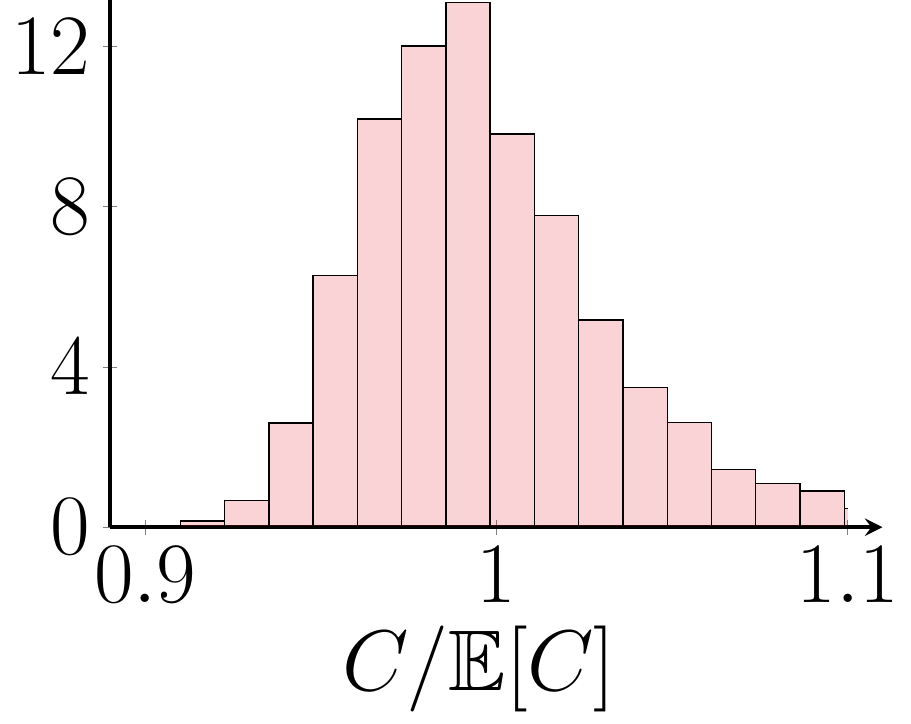}
		\caption{$n=10^4,\tau=2.5$}
		\label{fig:Chyp2510000}
	\end{subfigure}
	\begin{subfigure}{0.25\textwidth}
		\includegraphics[width=\textwidth]{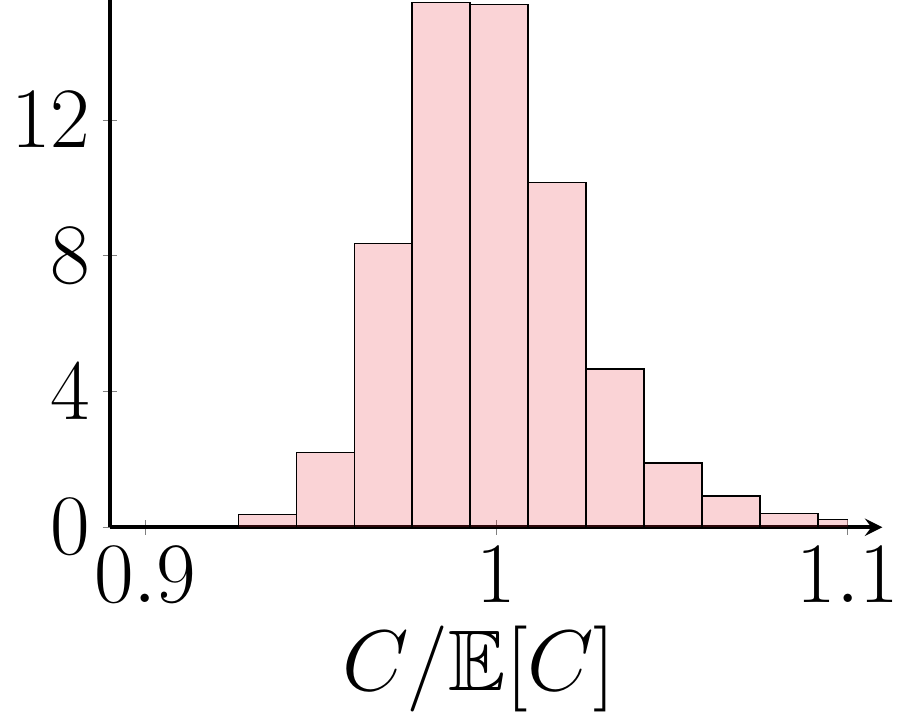}
		\caption{$n=10^4,\tau=2.8$}
		\label{fig:Chyp2810000}
	\end{subfigure}
	\\
	\begin{subfigure}{0.25\textwidth}
		\includegraphics[width=\textwidth]{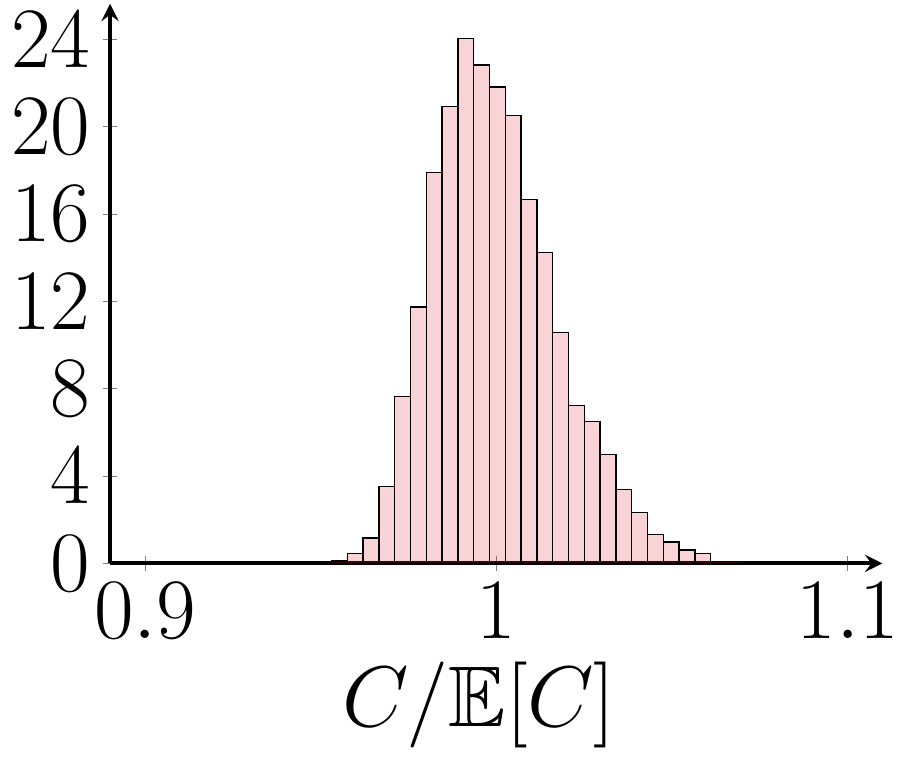}
		\caption{$n=10^5,\tau=2.2$}
		\label{fig:Chyp22100000}
	\end{subfigure}
	\hspace{0.1cm}
	\begin{subfigure}{0.25\textwidth}
		\includegraphics[width=\textwidth]{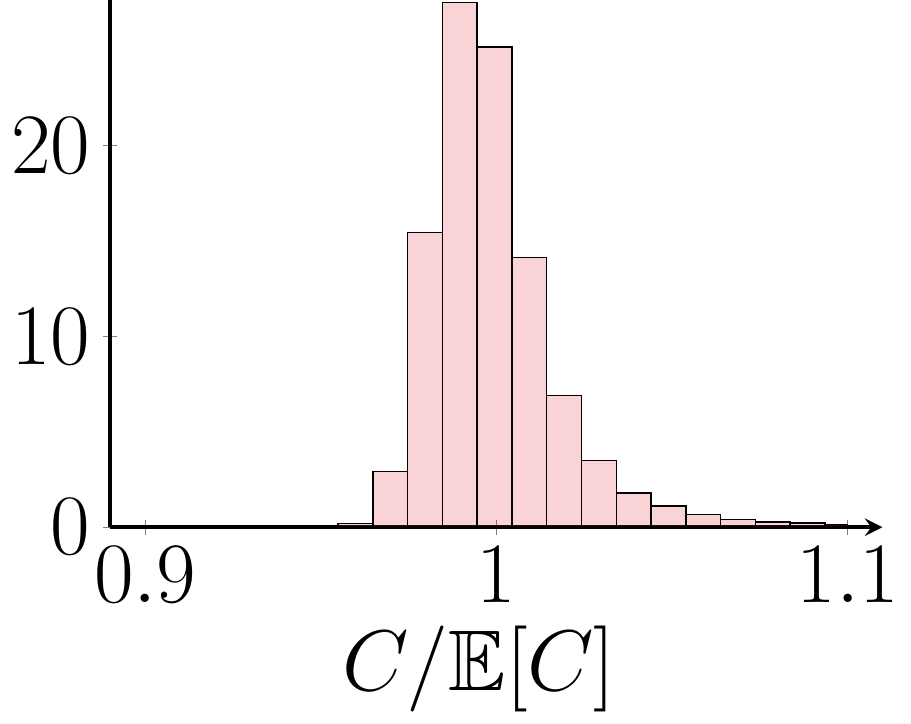}
		\caption{$n=10^5,\tau=2.5$}
		\label{fig:Chyp25100000}
	\end{subfigure}
	\begin{subfigure}{0.25\textwidth}
		\includegraphics[width=\textwidth]{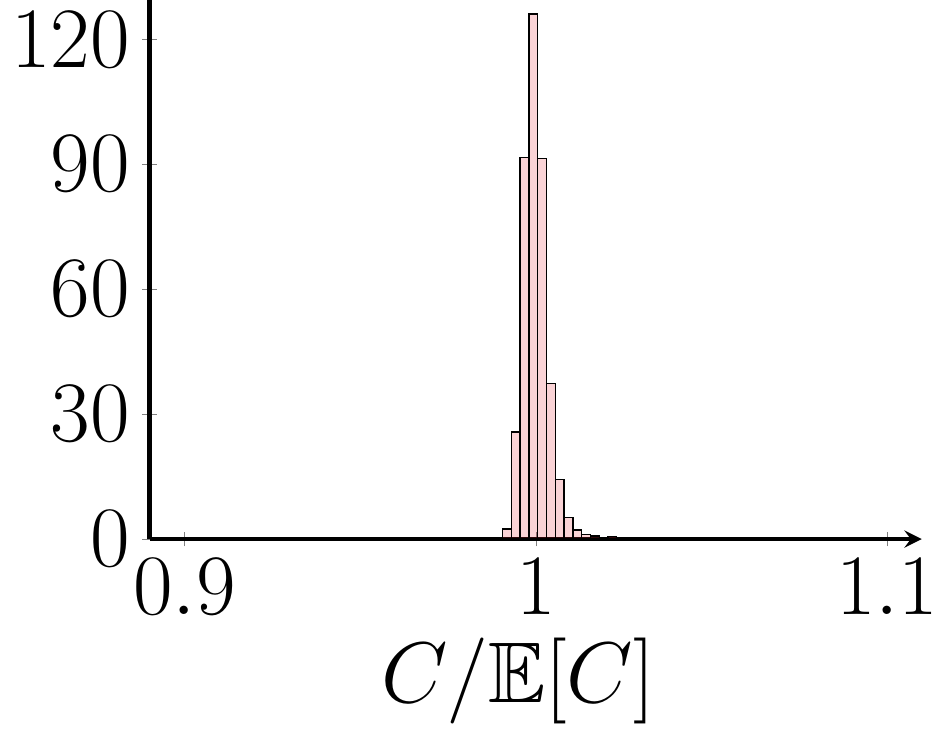}
		\caption{$n=10^5,\tau=2.8$}
		\label{fig:Chyp28100000}
	\end{subfigure}
	\caption{The self-averaging behavior of the clustering coefficient in the hyperbolic random graph. The plots show density estimates of the rescaled global clustering coefficient based on $10^4$ samples of hyperbolic random graphs with $\nu=1$.}
	\label{fig:selfavghyp}
\end{figure}

Instead of studying the local clustering curve $c(k)$, we now study the average clustering coefficient, defined as
\begin{equation}
C=\frac{1}{n}\sum_{i=1}^{n}\frac{N^\triangle_ i}{d_i(d_i-1)} =\sum_{k}p_kc(k),
\end{equation}
where $N^\triangle_i$ denotes the number of triangles attached to vertex $i$ and $p_k$ denotes the fraction of vertices of degree $k$. 
Because the power-law degree-distribution decays rapidly in $k$, $C\propto c(k)$ for constant $k$, since we know that $c(k)$ is approximately constant for constant $k$ (which was shown rigorously for the hidden-variable model~\cite{hofstad2017b}). Hence, the self-averaging properties of the average clustering coefficient are determined by the self-averaging properties of $c(k)$ for small values of $k$. Thus, in the hyperbolic random graph, the self-averaging $c(k)$-curve shows that also $C$ is self-averaging. Figure~\ref{fig:selfavghyp} shows that indeed the fluctuations in $C$ decrease as $n$ grows. 

\section{Other random graph models}\label{sec:treelike}
We next apply the variational principle~\eqref{eq:maxalphgen} to several random graph models. 
\subsection{Hidden-variable model}
\begin{figure*}[tb]
	\centering
	\begin{subfigure}[t]{0.2\linewidth}
		\centering
		\begin{tikzpicture}[treevertex/.style = {circle, align=center,red,fill=colorn1tau, draw=none,minimum size=0.4cm}, br/.style={draw=black!80,thick}]
		\tikzstyle{selected edge} = [draw,line width=5pt,-,black!40]
		[shorten >=1pt,auto,vertex distance=3cm,
		semithick]
		
		\node [treevertex] (z) at (0,0) {};
		\node [treevertex] (x) at (1,-1.2) {};
		\node [treevertex] (y) at (-1,-1.2) {};
		\draw[br] (z)--(y);
		\draw[br] (x)--(y);
		\draw[br] (z)--(x);
		
		\node (c) at ([shift={(100:0.3)}]z.60) {$k$};
		\node (d) at ([shift={(-25:0.4)}]x.60) {$d_1$};
		\node (e) at ([shift={(195:0.6)}]y.60) {$d_2$};
		\node (f) at (0,-1.6) {$d_1d_2\propto n$};
		\end{tikzpicture}
		\caption{$k\ll n^{\frac{\tau-2}{\tau-1}}$}
	\end{subfigure}
\hspace{0.8cm}
	\begin{subfigure}[t]{0.2\linewidth}
		\centering
	\begin{tikzpicture}[treevertex/.style = {circle, align=center,red,fill=colorn1tau, draw=none,minimum size=0.4cm}, br/.style={draw=black!80,thick}]
	\tikzstyle{selected edge} = [draw,line width=5pt,-,black!40]
	[shorten >=1pt,auto,vertex distance=3cm,
	semithick]
	
	\node [treevertex] (z) at (0,0) {};
	\node [treevertex] (x) at (1,-1.2) {};
	\node [treevertex] (y) at (-1,-1.2) {};
	\draw[br] (z)--(y);
	\draw[br] (x)--(y);
	\draw[br] (z)--(x);
	
	\node (c) at ([shift={(100:0.3)}]z.60) {$k$};
	\node (d) at ([shift={(-15:0.8)}]x.60) {$d_1<\tfrac{n}{k}$};
	\node (e) at ([shift={(190:0.9)}]y.60) {$d_2<\tfrac{n}{k}$};
	\node (f) at (0,-1.6) {$d_1d_2\propto n$};
	\end{tikzpicture}
	\caption{$ n^{\frac{\tau-2}{\tau-1}}\ll k\ll \sqrt{n}$}
\end{subfigure}
\hspace{1.8cm}
	\begin{subfigure}[t]{0.2\linewidth}
		\begin{tikzpicture}[treevertex/.style = {circle, align=center,red,fill=colorn1tau, draw=none,minimum size=0.4cm}, br/.style={draw=black!80,thick}]
		\tikzstyle{selected edge} = [draw,line width=5pt,-,black!40]
		[shorten >=1pt,auto,vertex distance=3cm,
		semithick]
		
		\node [treevertex] (z) at (0,0) {};
		\node [treevertex] (x) at (1,-1.2) {};
		\node [treevertex] (y) at (-1,-1.2) {};
		\draw[br] (z)--(y);
		\draw[br] (x)--(y);
		\draw[br] (z)--(x);
		
		\node (c) at ([shift={(100:0.3)}]z.60) {$k$};
		\node (d) at ([shift={(-25:0.4)}]x.60) {$\tfrac{n}{k}$};
		\node (e) at ([shift={(195:0.6)}]y.60) {$\tfrac{n}{k}$};
		\node (f) at (0,-1.6) {};
		\end{tikzpicture}
		\caption{$k\gg\sqrt{n}$}
	\end{subfigure}
	\caption{Typical triangles where one vertex has degree $k$ in the hidden-variable model. When $k<\sqrt{n}$ a typical triangle is with two vertices such that the product of their degrees is proportional to $n$. When $k>\sqrt{n}$, the other two degrees in a typical triangle are proportional to $n/k$.}
	\label{fig:ckhidden}
\end{figure*}
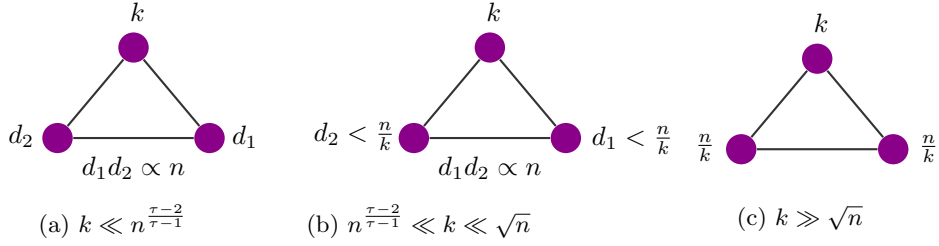

The hidden-variable model~\cite{boguna2003,chung2002} equips all vertices with a hidden variable $h$, an i.i.d.\ sample from a power-law distribution with degree exponent $\tau$. Vertices $i$ and $j$ with weights $h_i$ and $h_j$ connect with probability
\begin{equation}\label{eq:pij}
p(h_i,h_j)=\min(h_ih_j/(\mu n),1),
\end{equation}
where $\mu$ denotes the average weight.
Thus, the probability that a vertex of degree $k$ forms a triangle together with vertices $i$ and $j$ of degrees $n^{\alpha_1}$ and $n^{\alpha_2}$, respectively, can be written as
\begin{align}
\prob(\triangle_{i,j,k})=& \Theta\big(\min(kn^{\alpha_1-1},1)\min(kn^{\alpha_2-1},1)\min(n^{\alpha_1+\alpha_2-1},1)\big).
\end{align}
Therefore~\eqref{eq:maxalphgen} reduces to
\begin{align}\label{eq:ckalph}
\max_{\alpha_1,\alpha_2}&  \ n^{(\alpha_1+\alpha_2)(1-\tau)}\min(kn^{\alpha_1-1},1)\min(kn^{\alpha_2-1},1)\min(n^{\alpha_1+\alpha_2-1},1).
\end{align}
Calculating the optimum of~\eqref{eq:ckalph} over $\alpha_1,\alpha_2\in[0,1/(\tau-1)]$ shows that the maximal contribution to the typical number of constrained triangles is given by 
\begin{align}\label{eq:ckmaxcont}
&\alpha_1+\alpha_2=1, &&  k\ll n^{(\tau-2)/(\tau-1)},\nonumber\\
&\alpha_1+\alpha_2=1, n^{\alpha_1},n^{\alpha_2}<n/k, && n^{(\tau-2)/(\tau-1)}\ll k \ll \sqrt{n} ,\nonumber\\
&n^{\alpha_1}=n/k, n^{\alpha_2}=n/k, && k\gg \sqrt{n}.
\end{align}
Thus, for every value of $k$ there exists an optimal constrained triangle, visualized in Fig.~\ref{fig:ckhidden}. These three ranges of optimal triangle structures result in three ranges in $k$ for $c(k)$ in the hidden-variable model. Using these typical constrained subgraphs, we can characterize the entire spectrum of $c(k)$ as
\begin{equation}\label{eq:ckknownmain}
c(k)\propto\begin{cases}
n^{2-\tau}\log(n) & k\ll n^{(\tau-2)/(\tau-1)},\\
n^{2-\tau}\log(n/k^2) & n^{(\tau-2)/(\tau-1)}\ll k \ll \sqrt{n}, \\
k^{2\tau-6}n^{5-2\tau} & k\gg \sqrt{n}.
\end{cases}
\end{equation}
Figure~\ref{fig:ckhidden} shows that these three ranges are also visible in simulations.

The extended variational principle in Appendix~\ref{sec:fluct} shows that $c(k)$ in the hidden-variable model fails to be self-averaging for $k\ll n^{(\tau-2)/(\tau-1)}$, so that the values of $c(k)$ for $k\ll n^{(\tau-2)/(\tau-1)}$ heavily fluctuate across the various network samples. 

\subsection{Erased configuration model and uniform random graph}
The analysis of the optimal triangle structure in the hidden-variable model easily extends to two other important random graph models: the erased configuration model~\cite{britton2006}, where multiple edges and self-loops of the popular configuration model~\cite{bollobas1980} are removed, and the uniform random graph, a uniformly chosen graph from the ensemble of all simple graphs with a given degree distribution. Interestingly, both models can be approximated by a hidden-variable model with specific connection probabilities~\cite{hofstad2017c,gao2018}. 
The erased configuration model for example can be approximated by a hidden-variable model where the connection probabilities are given by~\cite{hofstad2017c}
\begin{equation}
p(h_i,h_j)=1-\me^{-h_ih_j/(\mu n)},
\end{equation}
and the uniform random graph can be approximated by a hidden-variable model with connection probabilities~\cite{gao2018}
\begin{equation}
p(h_i,h_j)=\frac{h_ih_j}{h_ih_j+\mu n}.
\end{equation}
Therefore, the optimal triangle structure as well as the behavior of the local clustering coefficient is the same as in~\eqref{eq:ckknownmain} and  Fig.~\ref{fig:ckhidden}. The non-self-averaging behavior for $k\ll n^{(\tau-2)/(\tau-1)}$ also extends from the hidden-variable model to the erased configuration model and the uniform random graph. 

\begin{figure}[htb]
	\centering
	\begin{subfigure}{0.25\textwidth}
		\includegraphics[width=\textwidth]{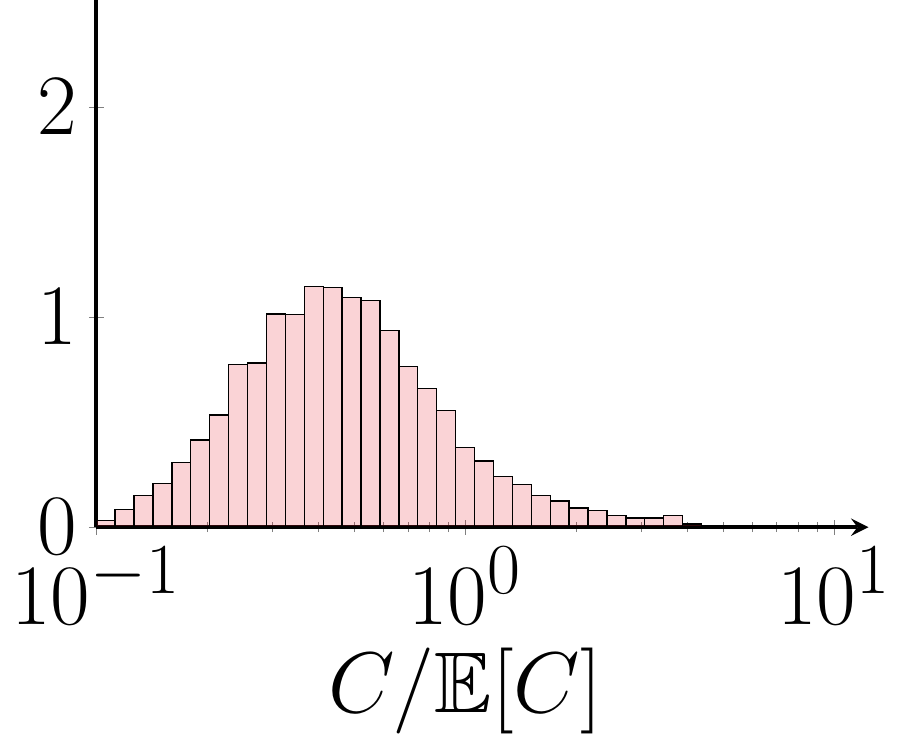}
		\caption{$n=10^4,\tau=2.5$}
		\label{fig:Chid2210000}
	\end{subfigure}
	\begin{subfigure}{0.25\textwidth}
		\includegraphics[width=\textwidth]{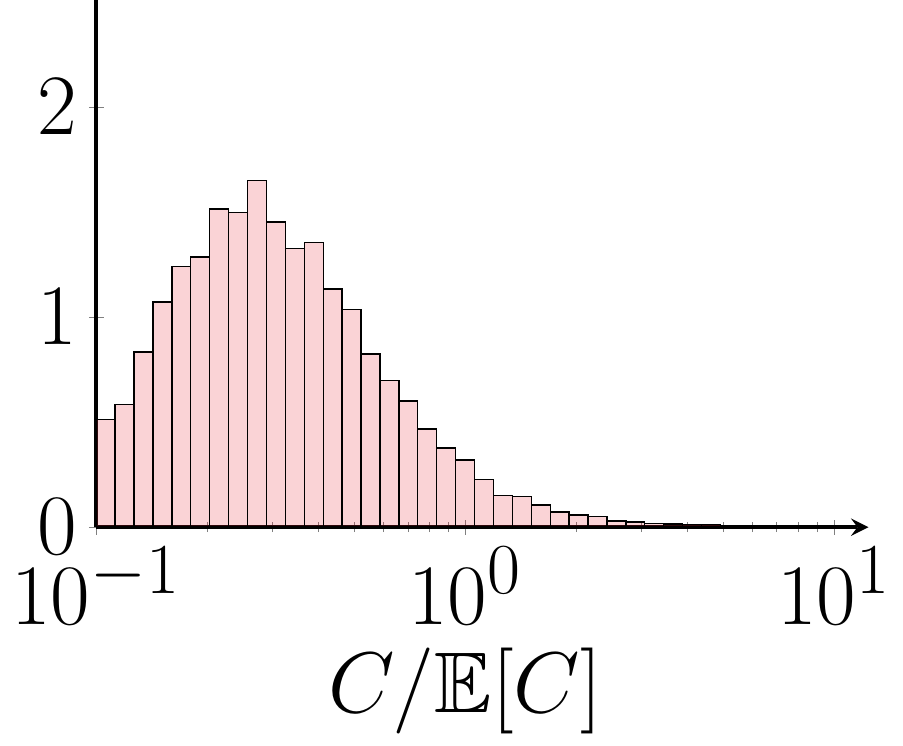}
		\caption{$n=10^4,\tau=2.5$}
		\label{fig:Chid2510000}
	\end{subfigure}
	\begin{subfigure}{0.25\textwidth}
		\includegraphics[width=\textwidth]{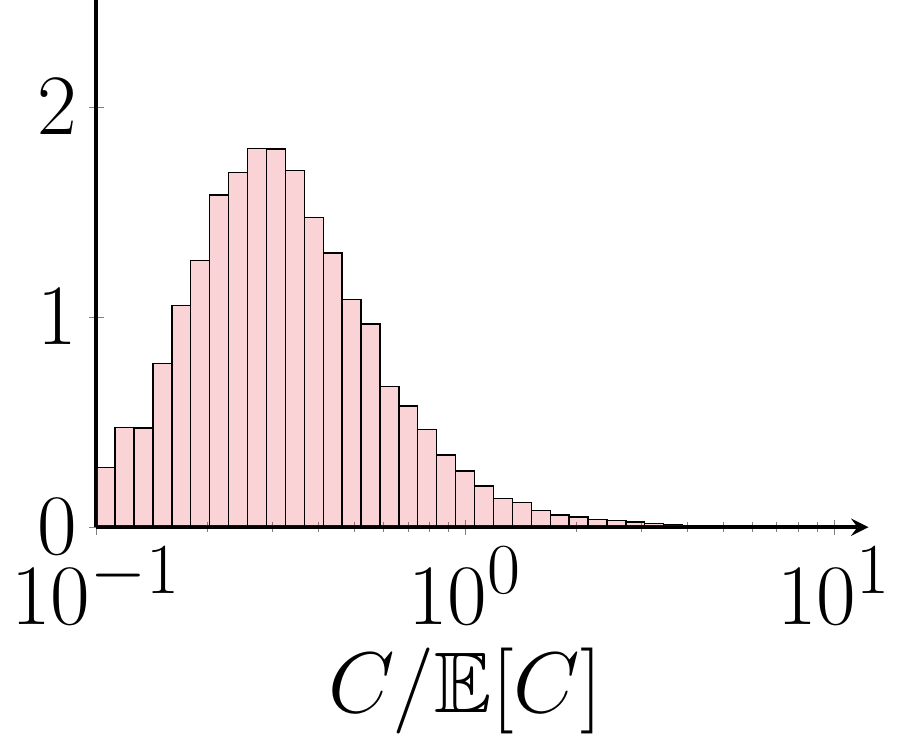}
		\caption{$n=10^4,\tau=2.8$}
		\label{fig:Chid2810000}
	\end{subfigure}
	\\
	\begin{subfigure}{0.25\textwidth}
		\includegraphics[width=\textwidth]{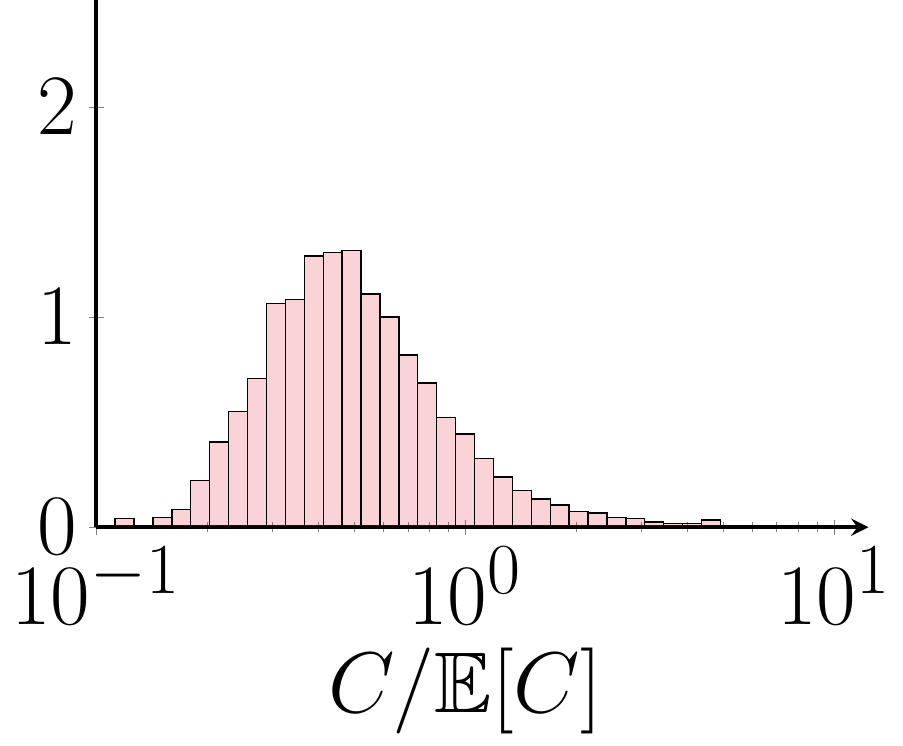}
		\caption{$n=10^5,\tau=2.5$}
		\label{fig:Chid22100000}
	\end{subfigure}
	\begin{subfigure}{0.25\textwidth}
		\includegraphics[width=\textwidth]{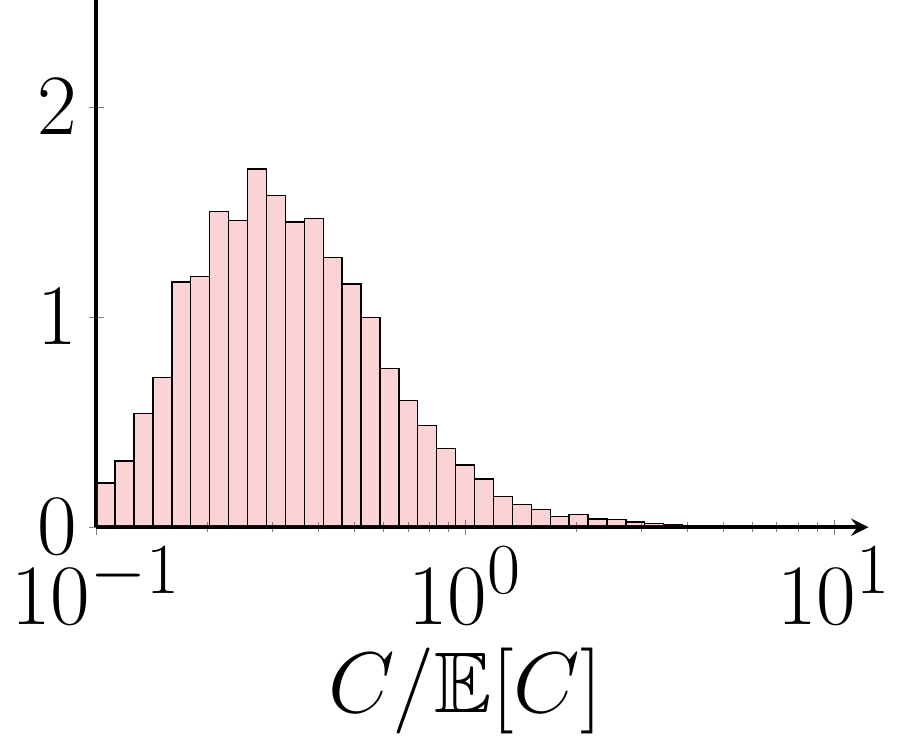}
		\caption{$n=10^5,\tau=2.5$}
		\label{fig:Chid25100000}
	\end{subfigure}
	\begin{subfigure}{0.25\textwidth}
		\includegraphics[width=\textwidth]{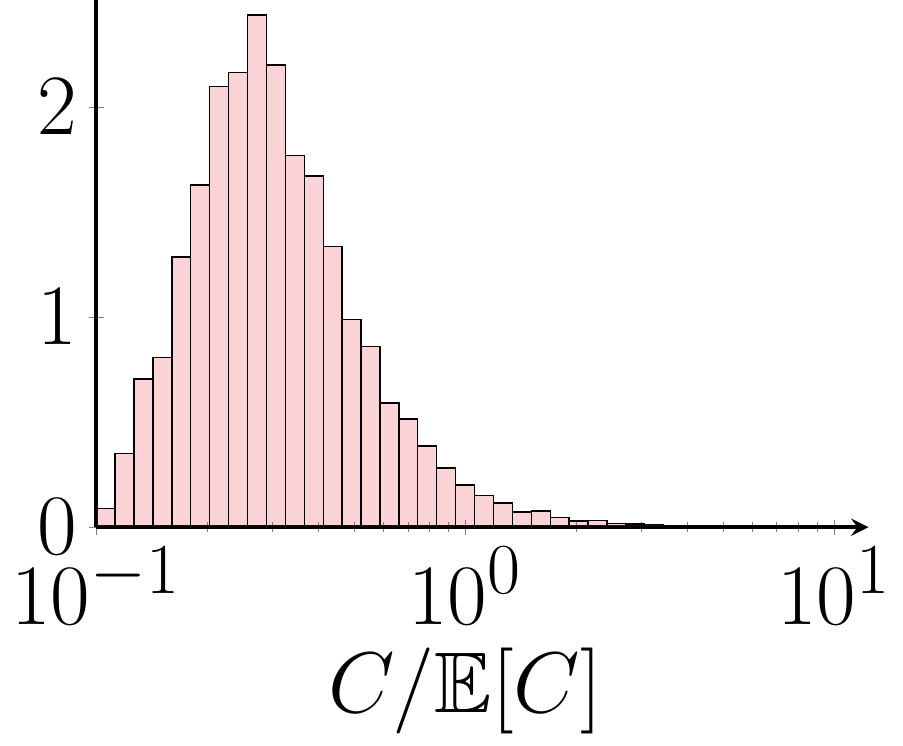}
		\caption{$n=10^5,\tau=2.8$}
		\label{fig:Chid28100000}
	\end{subfigure}
	\caption{The non-self-averaging behavior of the clustering coefficient in the hidden-variable model. The plots show density estimates based on $10^4$ samples of hidden-variable models.}
	\label{fig:selfavghid}
\end{figure}

In Appendix~\ref{sec:fluct} we show that $c(k)$ is non-self-averaging in the hidden-variable model when $k\ll n^{(\tau-2)/(\tau-1)}$. This also implies that the global clustering coefficient $C$ is non-self-averaging in the hidden-variable model, which supports numerical results in~\cite{colomer2012}.   Figure~\ref{fig:selfavghid} confirms that in the hidden-variable model $C$ is non-self-averaging, since the fluctuations in $C$ persist for large $n$.

\subsection{Preferential attachment}
Another important network null model is the preferential attachment model, a dynamic network model that can generate scale-free networks for appropriate parameter choices~\cite{albert1999a,buckley2004}. This model starts with two vertices, vertex 1 and 2, with $m$ edges between them. Then, at each step $t>2$, vertex $t$ is added with $m$ new edges attached to it. Each of these $m$ new edges attaches independently to an existing vertex $i<t$ with probability
\begin{equation}
\frac{d_i(t)+\delta}{2m t+\delta t},
\end{equation}
where $d_i(t)$ denotes the degree of vertex $i$ at time $t$.
This constructs a random graph with power-law degrees with exponent $\tau=3+\delta/m$. Thus, choosing $\delta\in(-m,0)$ constructs a random graph with exponent $\tau\in(2,3)$. 

In the preferential attachment model, it is convenient to apply the variational principle to vertices with index of a specific order of magnitude instead of degrees. The vertex with index 1 is the oldest vertex, and the vertex with index $n$ is the youngest vertex in the graph of size $n$. The probability that vertices with indices $i=n^{\alpha_1}$ and $j=n^{\alpha_2}$ with $\alpha_1<\alpha_2$ connect is proportional to~\cite{dommers2010}
\begin{equation}
\Prob{j \to i}\propto  j^{-\chi}i^{1-\chi}\propto n^{\alpha_1(\chi-1)-\alpha_2\chi},
\end{equation}
where $\chi=(\tau-2)/(\tau-1)$.  Thus, the probability that a vertex with fixed index $n^{\alpha_k}$ creates a triangle together with vertices of indices proportional to $n^{\alpha_1}$ and $n^{\alpha_2}$ can be approximated by
\begin{align}
& \Prob{\triangle \text{ on indices }n^{\alpha_k},n^{\alpha_1},n^{\alpha_2}}\nonumber\\
& \propto 
\begin{cases}
n^{2\alpha_1(\chi-1)-\alpha_2-2\alpha_k\chi} & \text{if }\alpha_1\leq \alpha_2\leq \alpha_k,\\
n^{2\alpha_1(\chi-1)-\alpha_k-2\alpha_2\chi} & \text{if }\alpha_1\leq \alpha_k\leq \alpha_2,\\
n^{2\alpha_k(\chi-1)-\alpha_1-2\alpha_2\chi} & \text{if }\alpha_k\leq \alpha_1\leq \alpha_2.
\end{cases}
\end{align}
The probability that a randomly chosen vertex has age proportional to $n^{\alpha}$ is proportional to $n^{\alpha-1}$. Thus, the equivalent optimization problem to~\eqref{eq:maxalphgen} becomes
\begin{equation}
\max_{\alpha_1\leq \alpha_2}\begin{cases}
n^{-2+\alpha_1(2\chi-1)-2\alpha_k\chi} & \text{if }\alpha_1\leq \alpha_2\leq \alpha_k,\\
n^{-2+(\alpha_1-\alpha_2)(2\chi-1)-\alpha_k} & \text{if }\alpha_1\leq \alpha_k\leq \alpha_2,\\
n^{-2+2\alpha_k(\chi-1)-\alpha_2(2\chi-1)} & \text{if }\alpha_k\leq \alpha_1\leq \alpha_2.
\end{cases}
\end{equation}
Using that $\chi\in(0,\tfrac{1}{2})$ when $\tau\in(2,3)$, we find that for all $0<\alpha_k<1$ the unique optimizer is obtained by $\alpha_1^*=0$ and $\alpha_2^*=1$. 
Furthermore, the degree of a vertex of index $i\propto n^{\alpha_i}$ at time $n$, $d_i(n)$ satisfies with high probability~\cite[Chapter 8]{hofstad2009}
\begin{equation}\label{eq:degindex}
d_i(n)\propto (n/i)^{1/(\tau-1)}\propto n^{\frac{1-\alpha_i}{\tau-1}}.
\end{equation}
Thus, vertices with age proportional to $n^{\alpha_1^*}$ have degrees proportional to $n^{1/(\tau-1)}$, whereas vertices with age proportional to $n^{\alpha_2^*}$ have degrees proportional to a constant. We conclude that for all $1\ll k \ll n^{1/(\tau-1)}$, in the most likely triangle containing a vertex of degree $k$ one of the other vertices has constant degree and the other has degree proportional to $n^{1/(\tau-1)}$. 

Similarly to~\eqref{eq:degindex}, a vertex of degree proportional to $n^{\gamma}$ has index proportional to $n^{1-\gamma(\tau-1)}$. Thus, when $k\propto n^{\gamma}$
\begin{equation}
c(n^{\gamma})\propto n^{2\gamma}n^2n^{-2-2\chi+1-1+\gamma(\tau-1)}= n^{\gamma(\tau-3)-2\chi}.
\end{equation}
Thus, for all $1\ll k\ll n^{1/(\tau-1)}$,
\begin{equation}\label{eq:ckpa}
c(k)\propto k^{\tau-3}n^{-2\chi}.
\end{equation}
Figure~\ref{fig:PAck} shows that this asymptotic slope in $k$ is a good fit in simulations.

\begin{figure}[tb]
	\centering
	\begin{minipage}[t]{0.5\textwidth}
		\centering
		\begin{tikzpicture}[treevertex/.style = {circle, align=center,red,fill=colorn1tau, draw=none,minimum size=0.4cm}, br/.style={draw=black!80,thick}]
		\tikzstyle{selected edge} = [draw,line width=5pt,-,black!40]
		\tikzstyle{selected edge2} = [draw,line width=2.55pt,-,black!40]
		[shorten >=1pt,auto,vertex distance=3cm,
		semithick]
		
		\node [treevertex] (z) at (0,0) {};
		\node [treevertex] (x) at (1.2,-1.5) {};
		\node [treevertex] (y) at (-1.2,-1.5) {};
		\draw[br] (z)--(y);
		\draw[br] (x)--(y);
		\draw[br] (z)--(x);
		
		\node (c) at ([shift={(100:0.3)}]z.60) {$k$};
		\node (d) at ([shift={(-25:0.4)}]x.60) {$1$};
		\node (e) at ([shift={(190:1)}]y.60) {$n^{1/(\tau-1)}$};
		\end{tikzpicture}
		\caption{The most likely triangle containing a vertex of degree $k$ in the preferential attachment model.}
		\label{fig:PAtriang}
	\end{minipage}
\end{figure}
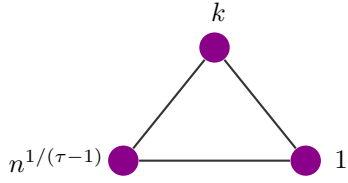

Figure~\ref{fig:PAtriang} shows the most likely triangle containing a vertex of degree $k$ in the preferential attachment model. Interestingly, this dominant triangle is the same over the entire range of $k$, which is very different from the three regimes that are present in the hidden-variable model.

\subsection{Random intersection graph}\label{sec:rig}
We next consider the random intersection graph~\cite{karonski1999}, a random graph model with overlapping community structures that, like the hyperbolic random graph, can generate non-vanishing clustering levels. The random intersection graph contains $n$ vertices, and $m$ vertex attributes. Every vertex $i$ chooses a random number of $X_i$ vertex attributes, where $(X_i)_{i\in[n]}$ is an i.i.d.\ sample. These vertex attributes are sampled uniformly without replacement from all $m$ attributes. Two vertices share an edge if they share at least $s\geq 1$ vertex attributes. One can think of the random intersection graph as a model for a social network, where every vertex attribute models the interest, or the group memberships of a person in the network. Then two vertices connect if their interests or group memberships are sufficiently similar. The overlapping community structures of the random intersection graph make the model highly clustered~\cite{bloznelis2013,bloznelis2013a}, so that the typical triangles in the random intersection graph should behave considerably different than the typical triangles in the locally tree-like models described above. 

To obtain random intersection graphs where vertices have asymptotically constant average degree, we need that $m^{s}\propto n$~\cite{bloznelis2013}, which we assume from now on. We further assume that $s$ is of constant order of magnitude. Then the degree of vertex $i$ with $X_i$ vertex attributes is proportional to $X_i^{s}$~\cite{bloznelis2013}. Therefore, a vertex of degree $k$ has approximately $k^{1/s}$ vertex attributes. To obtain a power-law degree distribution with exponent $\tau$, the probability of vertex $i$ having $X_i$ vertex attributes scales as
\begin{equation}
\Prob{X_i=u}\propto u^{-\tau s}.
\end{equation}

To apply the variational principle, we calculate the number of triangles between a vertex of degree $k$, and two vertices of degrees proportional to $n^{\alpha_1}$ and $n^{\alpha_2}$. These vertices have proportionally to $n^{\alpha_1/s}$, respectively $n^{\alpha_2/s}$, vertex attributes.
There are several ways for three vertices to form a triangle. If three vertices share the same set of at least $s$ attributes, then they form a triangle. But if vertex $i$ shares a set of at least $s$ attributes with vertex $j$, vertex $j$ shares another set of $s$ attributes with vertex $k$ and vertex $k$ shares yet another set of $s$ attributes with vertex $i$, these vertices also form a triangle. The most likely way for three vertices to form a triangle however, is for all three vertices to share the same set of $s$ attributes~\cite{bloznelis2013}. 
There are ${k^{1/s}\choose s}$ ways to choose $s$ attributes from the $k^{1/s}$ attributes of the degree-$k$ vertex. A triangle is formed if the two other vertices also contain these $s$ attributes. Since these vertices have $n^{\alpha_1/s}$ and $n^{\alpha_2/s}$ attributes chosen uniformly without replacement from all $m$ attributes, the probability that the first vertex shares these $s$ attributes is ${m-s\choose n^{\alpha_1/s}-s}/{m\choose n^{\alpha_1/s}}$. We then calculate the probability of a triangle being present as
\begin{align}
\Prob{\triangle \text{ on degrees }k,n^{\alpha_1},n^{\alpha_2}} & \propto
{k^{1/s}\choose s}\frac{{m-s\choose n^{\alpha_1/s}-s}{m-s\choose n^{\alpha_2/s}-s}}{{m\choose n^{\alpha_1/s}}{m\choose n^{\alpha_2/s}}}\nonumber\\
& \propto kn^{\alpha_1+\alpha_2}m^{-2s} \propto kn^{\alpha_1+\alpha_2-2}.
\end{align}
Combining this with~\eqref{eq:ckalphmax} yields
\begin{equation}
c(k)\propto n^2k^{-2}\max_{\alpha_1,\alpha_2}kn^{(\alpha_1+\alpha_2)(2-\tau)-2}\propto k^{-1},
\end{equation}
where the maximizer is $\alpha_1=\alpha_2=0$. Thus, a most likely triangle in the random intersection graph is a triangle containing one vertex of degree $k$, where the two other vertices have degrees proportional to a constant. 
The result that $c(k)\propto k^{-1}$ is in agreement with the results obtained in~\cite{bloznelis2013}. Moreover, the most likely triangle is a triangle where one vertex has degree $k$, and the other two vertices have constant degree. Thus, in terms of clustering, the random intersection graph behaves the same as the hyperbolic random graph with $\tau>5/2$.

\section{Discussion}

We have introduced a variational principle that finds the triangle that dominates clustering in scale-free random graph models. We have applied the variational principle to find optimal triangle structures in hidden-variable models, the preferential attachment model, random intersection graphs and the hyperbolic random graph, and believe that the variational principle can be applied to other random graph models such as the geometric inhomogeneous random graph~\cite{bringmann2015} or the spatial preferential attachment model~\cite{jacob2015,aiello2008}. We also presented an extended variational principle for general subgraphs to investigate the self-averaging properties of clustering. This method can also be applied to investigate higher order clustering~\cite{yin2018,benson2016}.

The hidden-variable model, erased configuration model, uniform random graph 
and preferential attachment model
all come with a clustering $c(k)$ that decreases with the network size.
This fall-off in $n$ can be understood in terms of the optimal triangle structures revealed by the variational principle. In all optimal triangle structures in Figs~\ref{fig:ckhidden} and~\ref{fig:PAtriang}, there is a vertex whose degree grows in $n$. 


In the hyperbolic model and the random intersection graph on the other hand, the optimal triangle structures in Fig.~\ref{fig:hyptyp} contain low-degree vertices for small values of $k$.
In models without geometric correlations, the probability of connecting two vertices usually increases with the degrees of the vertices. Therefore, models without correlations mostly contain triangles with high-degree vertices, causing these networks to be locally tree-like. The geometric correlations in the hyperbolic model on the other hand make it more likely for two low-degree neighbors to connect, causing the most likely triangle to contain lower-degree vertices. Lower-degree vertices are abundant, which explains why for small $k$, $c(k)$ does not vanish as $n$ grows large in the hyperbolic model, which is also observed in many real-world networks. 

Another advantage of the hyperbolic random graph over the locally tree-like networks is that the hyperbolic model is self-averaging over the entire range of $k$. 
This makes the local clustering curve more stable in the sense that it suffices to generate one large hyperbolic random graph to investigate the behavior of $c(k)$. 

\paragraph{Acknowledgements.}
This work is supported by NWO TOP grant 613.001.451 and by the NWO Gravitation Networks grant 024.002.003.

	\bibliographystyle{abbrv}
\bibliography{../references}

\appendix

\tikzstyle{every node}=[circle,fill=black!25,minimum size=8pt,inner sep=0pt,draw=black!80]

\section{Fluctuations in the hidden-variable model}\label{sec:fluct}
As in the hyperbolic random graph, we first study $\Exp{c(k)}$ by relaxing the constraint $\alpha\in[0,1/(\tau-1)]$ in~\eqref{eq:ckalph}.
As long as $k\gg n^{(\tau-2)/(\tau-1)}$, we see from~\eqref{eq:ckmaxcont} that the largest contribution to $c(k)$ is from vertices with degrees strictly smaller than $n^{1/(\tau-1)}$. Thus, removing the constraint on the maximal degree does not influence the major contribution for $c(k)$. When $k\ll n^{(\tau-2)/(\tau-1)}$ however, the major contribution includes vertices of degree $n^{1/(\tau-1)}$. Removing the constraint then results in an optimal contribution which is slightly different from~\eqref{eq:ckmaxcont}:
\begin{equation}\label{eq:ckmaxcontexp}
\begin{aligned}
&\alpha_1+\alpha_2=1, n^{\alpha_1},n^{\alpha_2}<n/k &&  k \ll \sqrt{n}, \\
&n^{\alpha_1}=n/k, n^{\alpha_2}=n/k && k\gg \sqrt{n}.
\end{aligned}
\end{equation}
Similarly to the computation that leads to~\eqref{eq:ckmaxcont}, this gives for $\Exp{c(k)}$ that
\begin{equation}\label{ckexp}
\Exp{c(k)}\propto\begin{cases}
n^{2-\tau}\log(n/k^2) & k \ll \sqrt{n}, \\
k^{2\tau-6}n^{5-2\tau} & k\gg \sqrt{n}.
\end{cases}
\end{equation}
Thus, the typical behavior of $c(k)$ is the same as its average behavior for $k\gg n^{(\tau-2)/(\tau-1)}$. For small values of $k$ however, the flat regime disappears and is replaced by a regime that depends on the logarithm of $k$. 

We now compute the variance of $c(k)$, again using~\eqref{eq:varconst}. We first investigate $\Exp{\scalebox{0.8}{\bowa}}=n^3N_k^2\Prob{\scalebox{0.8}{\bowa}}$
where $\prob\big(\scalebox{0.8}{\bowa}\big)$ denotes the probability that two randomly chosen vertices of degree $k$ form the constrained bow-tie together with three randomly chosen other vertices. As for the hyperbolic model, we compute this probability with a constrained variational principle. By symmetry of the bow-tie subgraph, the optimal degree range of the bottom right vertex and the upper right vertex is the same. Let the degree of the middle vertex scale as $n^{\alpha_1}$, and the degrees of the other two vertices as $n^{\alpha_2}$. Then, we write the constrained variational principle, similarly to~\eqref{eq:ckalph}, as
\begin{align}
\max_{\alpha_1,\alpha_2} & n^{(\alpha_1+2\alpha_2)(1-\tau)}\min(kn^{\alpha_1-1},1)^2\min(kn^{\alpha_2-1},1)^2\nonumber\\
& \times \min(n^{\alpha_1+\alpha_2-1},1)^2
\end{align}

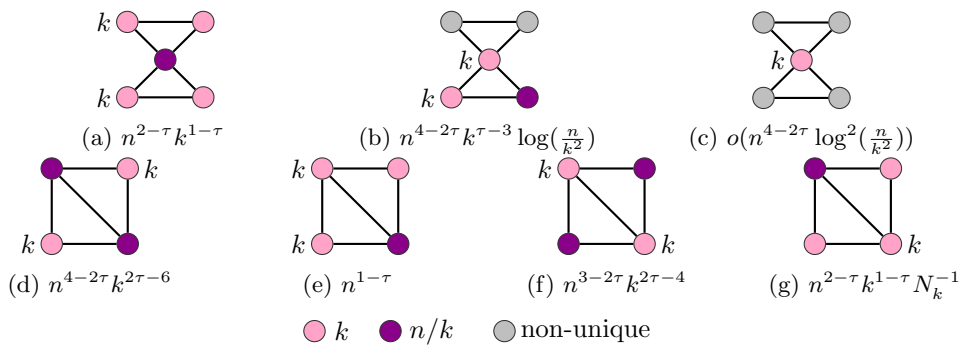
\begin{figure}[htb]
	\centering
	\begin{subfigure}[t]{0.25\linewidth}
		\centering
		\begin{tikzpicture}
		\centering
		\tikzstyle{edge} = [draw,thick,-]
		\node[S1,label=left:{$k$}] (a) at (0,0) {};
		\node[S1] (b) at (1,0) {};
		\node[S1,label=left:{$k$}] (c) at (0,1) {};
		\node[S1] (d) at (1,1) {};
		\node[S2m] (e) at (0.5,0.5) {};
		\draw[edge] (a)--(b);
		\draw[edge] (e)--(a);
		\draw[edge] (b)--(e);
		\draw[edge] (e)--(d);
		\draw[edge] (c)--(e);
		\draw[edge] (c)--(d);
		\end{tikzpicture}
		\caption{$n^{2-\tau}k^{1-\tau}$}
		\label{fig:bowk1}
	\end{subfigure}	
\hspace{0.3cm}
	\begin{subfigure}[t]{0.25\linewidth}
		\centering
		\begin{tikzpicture}
		\tikzstyle{edge} = [draw,thick,-]
		\node[S1,label=left:{$k$}] (a) at (0,0) {};
		\node[S2m] (b) at (1,0) {};
		\node (c) at (0,1) {};
		\node (d) at (1,1) {};
		\node[S1,label=left:{$k$}] (e) at (0.5,0.5) {};		\draw[edge] (a)--(b);
		\draw[edge] (e)--(a);
		\draw[edge] (b)--(e);
		\draw[edge] (e)--(d);
		\draw[edge] (c)--(e);
		\draw[edge] (c)--(d);
		\end{tikzpicture}
		\caption{$n^{4-2\tau}k^{\tau-3}\log(\frac{n}{k^2})$}
		\label{fig:bowk2}
	\end{subfigure}	
\hspace{0.3cm}
	\begin{subfigure}[t]{0.25\linewidth}
		\centering
		\begin{tikzpicture}
		\tikzstyle{edge} = [draw,thick,-]
		\node[] (a) at (0,0) {};
		\node[] (b) at (1,0) {};
		\node[] (c) at (0,1) {};
		\node[] (d) at (1,1) {};
		\node[S1,label=left:{$k$}] (e) at (0.5,0.5) {};
		\draw[edge] (a)--(b);
		\draw[edge] (e)--(a);
		\draw[edge] (b)--(e);
		\draw[edge] (e)--(d);
		\draw[edge] (c)--(e);
		\draw[edge] (c)--(d);
		\end{tikzpicture}
		\caption{$o(n^{4-2\tau}\log^2(\frac{n}{k^2}))$}
		\label{fig:bowk3}
	\end{subfigure}	
	
	\begin{subfigure}[t]{0.22\linewidth}
		\centering
		\begin{tikzpicture}
		\tikzstyle{edge} = [draw,thick,-]
		\node[S1,label=left:{$k$}] (a) at (0,0) {};
		\node[S2m] (b) at (1,0) {};
		\node[S2m] (c) at (0,1) {};
		\node[S1,label=right:{$k$}] (d) at (1,1) {};
		\draw[edge] (a)--(b);
		\draw[edge] (c)--(b);
		\draw[edge] (d)--(b);
		\draw[edge] (a)--(c);
		\draw[edge] (c)--(d);
		\end{tikzpicture}	
		\caption{$n^{4-2\tau}k^{2\tau-6}$}
		\label{fig:diamondk1}
	\end{subfigure}
	\begin{subfigure}[t]{0.22\linewidth}
		\centering
		\begin{tikzpicture}
		\tikzstyle{edge} = [draw,thick,-]
		\node[S1,label=left:{$k$}] (a) at (0,0) {};
		\node[S2m] (b) at (1,0) {};
		\node[S1,label=left:{$k$}] (c) at (0,1) {};
		\node[S1] at (1,1) {};
		\draw[edge] (a)--(b);
		\draw[edge] (c)--(b);
		\draw[edge] (d)--(b);
		\draw[edge] (a)--(c);
		\draw[edge] (c)--(d);
		\end{tikzpicture}	
		\caption{$n^{1-\tau}$}
		\label{fig:diamondk2}
	\end{subfigure}
	\begin{subfigure}[t]{0.22\linewidth}
		\centering
		\begin{tikzpicture}
		\tikzstyle{edge} = [draw,thick,-]
		\node[S2m] (a) at (0,0) {};
		\node[S1,label=right:{$k$}] (b) at (1,0) {};
		\node[S1,label=left:{$k$}] (c) at (0,1) {};
		\node[S2m] at (1,1) {};
		\draw[edge] (a)--(b);
		\draw[edge] (c)--(b);
		\draw[edge] (d)--(b);
		\draw[edge] (a)--(c);
		\draw[edge] (c)--(d);
		\end{tikzpicture}	
		\caption{$n^{3-2\tau}k^{2\tau-4}$}
		\label{fig:diamondk3}
	\end{subfigure}
	\begin{subfigure}[t]{0.22\linewidth}
		\centering
		\begin{tikzpicture}
		\tikzstyle{edge} = [draw,thick,-]
		\node[S1] (a) at (0,0) {};
		\node[S1,label=right:{$k$}](b) at (1,0) {};
		\node[S2m] (c) at (0,1) {};
		\node[S1] (d) at (1,1) {};
		\draw[edge] (a)--(b);
		\draw[edge] (c)--(b);
		\draw[edge] (d)--(b);
		\draw[edge] (a)--(c);
		\draw[edge] (c)--(d);
		\end{tikzpicture}	
		\caption{$n^{2-\tau}k^{1-\tau}N_k^{-1}$}
		\label{fig:diamondk4}
	\end{subfigure}
	
	\vspace{-0.5cm}
	\begin{subfigure}{\linewidth}
		\centering
		\begin{tikzpicture}
		\node[S2m,label={[label distance=0.05cm]0:$n/k$}] (a) at (2,0) {};
		\node[S1,label={[label distance=0.05cm]0:$k$}] (c) at (1,0) {};
		\node[label={[label distance=0.05cm]0:non-unique}] (d) at (3.5,0) {};
		\end{tikzpicture}
	\end{subfigure}
	\vspace{-0.8cm}
	
	\caption{Contribution to the variance of $c(k)$ in the hidden-variable model from merging two triangles where one vertex has degree $k\ll \sqrt{n}$. The vertex color indicates the optimal vertex degree.}
	\label{fig:bowtiek}
\end{figure}

We then find that for $k\ll\sqrt{n}$, the unique optimal contribution is from $n^{\alpha_1}=n/k$ and $n^{\alpha_2}=k$, as shown in Fig.~\ref{fig:bowk1}. 
Thus, the expected number of such bow-ties scales as
\begin{equation}
\Exp{\bowa}\propto n^3N_k^2(n/k)^{1-\tau}k^{2(1-\tau)}k^4n^{-2}=N_k^2n^{2-\tau}k^{5-\tau}.
\end{equation}
Thus,
\begin{equation}\label{eq:varckbound}
\Var{c(k)}> k^{-4}N_k^{-2}\Exp{\bowa}\propto n^{2-\tau}k^{1-\tau},
\end{equation}
so that~\eqref{ckexp} yields that for $k$ small
\begin{equation}
\frac{\Var{c(k)}}{\Exp{c(k)}^2}>\frac{n^{2-\tau} k^{1-\tau}}{n^{4-2\tau}\log^2(n/k^2)},
\end{equation}
which tends to infinity as long as $k\ll n^{(\tau-2)/(\tau-1)}$. Therefore $c(k)$ is non self-averaging as long as $k\ll n^{(\tau-2)/(\tau-1)}$. 

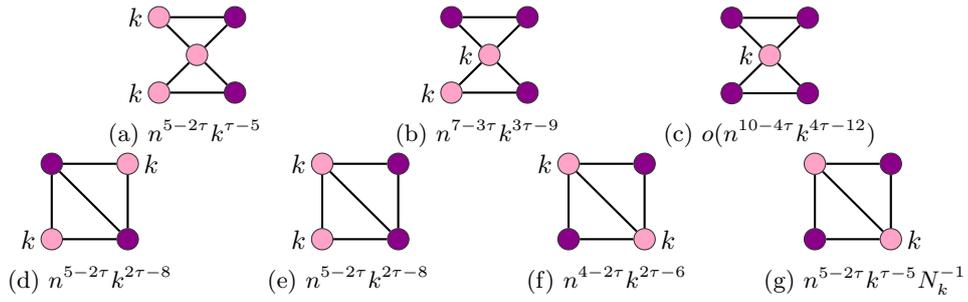
\begin{figure}[htb]
	\centering
	\begin{subfigure}[t]{0.25\linewidth}
		\centering
		\begin{tikzpicture}
		\centering
		\tikzstyle{edge} = [draw,thick,-]
		\node[S1,label=left:{$k$}] (a) at (0,0) {};
		\node[S2m] (b) at (1,0) {};
		\node[S1,label=left:{$k$}] (c) at (0,1) {};
		\node[S2m] (d) at (1,1) {};
		\node[S1] (e) at (0.5,0.5) {};
		\draw[edge] (a)--(b);
		\draw[edge] (e)--(a);
		\draw[edge] (b)--(e);
		\draw[edge] (e)--(d);
		\draw[edge] (c)--(e);
		\draw[edge] (c)--(d);
		\end{tikzpicture}
		\caption{$n^{5-2\tau}k^{\tau-5}$}
		\label{fig:bowk1s}
	\end{subfigure}	
	\begin{subfigure}[t]{0.25\linewidth}
		\centering
		\begin{tikzpicture}
		\tikzstyle{edge} = [draw,thick,-]
		\node[S1,label=left:{$k$}] (a) at (0,0) {};
		\node[S2m] (b) at (1,0) {};
		\node[S2m] (c) at (0,1) {};
		\node[S2m] (d) at (1,1) {};
		\node[S1,label=left:{$k$}] (e) at (0.5,0.5) {};
		\draw[edge] (a)--(b);
		\draw[edge] (e)--(a);
		\draw[edge] (b)--(e);
		\draw[edge] (e)--(d);
		\draw[edge] (c)--(e);
		\draw[edge] (c)--(d);
		\end{tikzpicture}
		\caption{$n^{7-3\tau}k^{3\tau-9}$}
		\label{fig:bowk2s}
	\end{subfigure}	
	\begin{subfigure}[t]{0.25\linewidth}
		\centering
		\begin{tikzpicture}
		\tikzstyle{edge} = [draw,thick,-]
		\node[S2m] (a) at (0,0) {};
		\node[S2m] (b) at (1,0) {};
		\node[S2m] (c) at (0,1) {};
		\node[S2m] (d) at (1,1) {};
		\node[S1,label=left:{$k$}] (e) at (0.5,0.5) {};
		\draw[edge] (a)--(b);
		\draw[edge] (e)--(a);
		\draw[edge] (b)--(e);
		\draw[edge] (e)--(d);
		\draw[edge] (c)--(e);
		\draw[edge] (c)--(d);
		\end{tikzpicture}
		\caption{$o(n^{10-4\tau}k^{4\tau-12})$}
		\label{fig:bowk3s}
	\end{subfigure}	
	
	\begin{subfigure}[t]{0.22\linewidth}
		\centering
		\begin{tikzpicture}
		\tikzstyle{edge} = [draw,thick,-]
		\node[S1,label=left:{$k$}] (a) at (0,0) {};
		\node[S2m] (b) at (1,0) {};
		\node[S2m] (c) at (0,1) {};
		\node[S1,label=right:{$k$}] (d) at (1,1) {};
		\draw[edge] (a)--(b);
		\draw[edge] (c)--(b);
		\draw[edge] (d)--(b);
		\draw[edge] (a)--(c);
		\draw[edge] (c)--(d);
		\end{tikzpicture}	
		\caption{$n^{5-2\tau}k^{2\tau-8}$}
		\label{fig:diamondk1s}
	\end{subfigure}
	\begin{subfigure}[t]{0.22\linewidth}
		\centering
		\begin{tikzpicture}
		\tikzstyle{edge} = [draw,thick,-]
		\node[S1,label=left:{$k$}] (a) at (0,0) {};
		\node[S2m] (b) at (1,0) {};
		\node[S1,label=left:{$k$}] (c) at (0,1) {};
		\node[S2m] at (1,1) {};
		\draw[edge] (a)--(b);
		\draw[edge] (c)--(b);
		\draw[edge] (d)--(b);
		\draw[edge] (a)--(c);
		\draw[edge] (c)--(d);
		\end{tikzpicture}	
		\caption{$n^{5-2\tau}k^{2\tau-8}$}
		\label{fig:diamondk2s}
	\end{subfigure}
	\begin{subfigure}[t]{0.22\linewidth}
		\centering
		\begin{tikzpicture}
		\tikzstyle{edge} = [draw,thick,-]
		\node[S2m] (a) at (0,0) {};
		\node[S1,label=right:{$k$}] (b) at (1,0) {};
		\node[S1,label=left:{$k$}] (c) at (0,1) {};
		\node[S2m] at (1,1) {};
		\draw[edge] (a)--(b);
		\draw[edge] (c)--(b);
		\draw[edge] (d)--(b);
		\draw[edge] (a)--(c);
		\draw[edge] (c)--(d);
		\end{tikzpicture}	
		\caption{$n^{4-2\tau}k^{2\tau-6}$}
		\label{fig:diamondk3s}
	\end{subfigure}
	\begin{subfigure}[t]{0.22\linewidth}
		\centering
		\begin{tikzpicture}
		\tikzstyle{edge} = [draw,thick,-]
		\node[S2m] (a) at (0,0) {};
		\node[S1,label=right:{$k$}](b) at (1,0) {};
		\node[S1] (c) at (0,1) {};
		\node[S2m] (d) at (1,1) {};
		\draw[edge] (a)--(b);
		\draw[edge] (c)--(b);
		\draw[edge] (d)--(b);
		\draw[edge] (a)--(c);
		\draw[edge] (c)--(d);
		\end{tikzpicture}	
		\caption{$n^{5-2\tau}k^{\tau-5}N_k^{-1}$}
		\label{fig:diamondk4s}
	\end{subfigure}
	\caption{Contribution to the variance of $c(k)$ in the hidden-variable model from merging two triangles where one vertex has degree $k\gg \sqrt{n}$. The vertex color indicates the optimal vertex degree as in Fig.~\ref{fig:bowtiek}.}
	\label{fig:bowtieks}
\end{figure}

For $n^{(\tau-2)/(\tau-1)}\ll k \ll \sqrt{n}$, we can similarly compute the optimum contributions of all other constrained motifs to the variance as in Fig.~\ref{fig:bowtiek}. This shows that $c(k)$ is self-averaging, since all contributions have smaller magnitude than $\Exp{c(k)}^2$ (obtained from~\eqref{ckexp}). For $k\gg\sqrt{n}$, a constrained variational principle again provides the contribution of all constrained motifs to the variance of $c(k)$, as visualized in Fig.~\ref{fig:bowtieks}. Comparing this with~\eqref{ckexp} shows that $c(k)$ is also self-averaging for $k\gg\sqrt{n}$. 

\end{document}